\documentclass[journal]{IEEEtran}
\ifCLASSINFOpdf
\else
\fi
\usepackage{url}

\usepackage{mathtools,booktabs}
\usepackage{multirow}
\usepackage{nicefrac}       
\usepackage{amsfonts,amsthm,pifont}
\usepackage{algpseudocode}
\usepackage{cite,bm,amsmath,amssymb,graphicx,subfigure,enumitem,float,xfrac,mathtools}
\usepackage[mathscr]{euscript}

\usepackage{hyperref}
\hypersetup{colorlinks=black, pdfstartview=FitV, linkcolor=blue, citecolor=black, plainpages=false, pdfpagelabels=true, urlcolor=blue}
\usepackage[all]{hypcap}

\usepackage{relsize}

\usepackage{caption}
\usepackage{color}
\usepackage{dcolumn}

\allowdisplaybreaks

\usepackage{etoolbox}
\makeatletter
\patchcmd{\maketitle}
 {\def\@makefnmark}
 {\def\@makefnmark{}\def\useless@macro}
 {}{}
\makeatother

\usepackage{tikz}

\newcommand{\vw}{{\boldsymbol w}}
\newcommand{\vx}{{\boldsymbol x}}
\newcommand{\vy}{{\boldsymbol y}}
\newcommand{\vz}{{\boldsymbol z}}

\newcommand{\vr}{{\boldsymbol r}}
\newcommand{\vs}{{\boldsymbol s}}
\newcommand{\vu}{{\boldsymbol u}}
\newcommand{\vv}{{\boldsymbol v}}

\newcommand{\vA}{{\boldsymbol A}}
\newcommand{\vX}{{\boldsymbol X}}

\newcommand{\vU}{{\boldsymbol U}}
\newcommand{\vD}{{\boldsymbol D}}
\newcommand{\vF}{{\boldsymbol F}}
\newcommand{\vR}{{\boldsymbol R}}

\newcommand{\vI}{{\boldsymbol I}}

\usepackage{chngcntr}
\usepackage{apptools}
\AtAppendix{\counterwithin{proposition}{section}}

\newtheorem{proposition}{Proposition}

\newtheorem*{remark}{Remark}

\newenvironment{sef_ene}[1][\ding{113} SEF:\\]{\begin{trivlist}
\item[\hskip \labelsep {\bfseries #1}]}{\end{trivlist}}

\newenvironment{ref_ene}[1][\ding{113} REF:\\]{\begin{trivlist}
\item[\hskip \labelsep {\bfseries #1}]}{\end{trivlist}}

\newenvironment{noiseless_r}[1][\ding{113} Noiseless recovery:\\]{\begin{trivlist}
\item[\hskip \labelsep {\bfseries #1}]}{\end{trivlist}}

\newenvironment{noisy_r}[1][\ding{113} Noisy recovery:]{\begin{trivlist}
\item[\hskip \labelsep {\bfseries #1}]}{\end{trivlist}}

\newenvironment{local_minimums}[1][\ding{113} Local minimums:\\]{\begin{trivlist}
\item[\hskip \labelsep {\bfseries #1}]}{\end{trivlist}}

\newenvironment{first_step}[1][Step 1:\\]{\begin{trivlist}
\item[\hskip \labelsep {\bfseries #1}]}{\end{trivlist}}

\newenvironment{second_step}[1][Step 2:\\]{\begin{trivlist}
\item[\hskip \labelsep {\bfseries #1}]}{\end{trivlist}}

\newenvironment{positive_threshold}[1][\ding{113} Positive threshold:\\]{\begin{trivlist}
\item[\hskip \labelsep {\bfseries #1}]}{\end{trivlist}}

\newenvironment{negative_threshold}[1][\ding{113} Negative threshold:\\]{\begin{trivlist}
\item[\hskip \labelsep {\bfseries #1}]}{\end{trivlist}}

\usepackage{algorithm,algcompatible}

\begin{document}
\onecolumn
\textcopyright 2019 IEEE.  Personal use of this material is permitted.  Permission from IEEE must be obtained for all other uses, in any current or future media, including reprinting/republishing this material for advertising or promotional purposes, creating new collective works, for resale or redistribution to servers or lists, or reuse of any copyrighted component of this work in other works.
\twocolumn
\newpage
\setcounter{page}{1}
\title{Sparse Signal Recovery via Generalized Entropy Functions Minimization}
%
%
%

\author{Shuai Huang, and Trac D. Tran, ~\IEEEmembership{Fellow,~IEEE}
\thanks{This work is supported by the National Science Foundation under grants NSF-CCF-1117545, NSF-CCF-1422995 and NSF-EECS-1443936.}
\thanks{The authors are with the Department of Electrical and Computer Engineering, Johns Hopkins University, Baltimore, MD, 21218 USA (email: shuaihuang@jhu.edu; trac@jhu.edu).}}

\maketitle

\begin{abstract}
Compressive sensing relies on the sparse prior imposed on the signal of interest to solve the ill-posed recovery problem in an under-determined linear system. The objective function used to enforce the sparse prior information should be both effective and easily optimizable. Motivated by the entropy concept from information theory, in this paper we propose the generalized Shannon entropy function and R\'{e}nyi entropy function of the signal as the sparsity promoting regularizers. Both entropy functions are nonconvex, non-separable. Their local minimums only occur on the boundaries of the orthants in the Euclidean space. Compared to other popular objective functions, minimizing the generalized entropy functions adaptively promotes multiple high-energy coefficients while suppressing the rest low-energy coefficients. The corresponding optimization problems can be recasted into a series of reweighted $l_1$-norm minimization problems and then solved efficiently by adapting the FISTA. Sparse signal recovery experiments on both the simulated and real data show the proposed entropy functions minimization approaches perform better than other popular approaches and achieve state-of-the-art performances. 
\end{abstract}

\begin{IEEEkeywords}
Compressive sensing, entropy functions minimization, image recovery, sparse representation classification
\end{IEEEkeywords}

%
\IEEEpeerreviewmaketitle

\section{Introduction}
Nowadays, there is an increasing amount of digital information generated constantly from every aspect of our life: the data that we work with grow in both size and variety. Fortunately, most of the data have inherent sparse structures. Compressive sensing \cite{CS06,IntroCS06,IntroCS08} offers us an efficient framework to not only collect those data, but also process and analyze them in a timely fashion. Various compressive sensing tasks eventually boil down to the sparse signal recovery problem in an under-determined linear system:
\begin{align}
\vy=\vA\vx+\vw\,,
\end{align}
where $\vy\in\mathbb{R}^M$ is the linear measurement, $\vA\in\mathbb{R}^{M\times N}$ is the sensing matrix, $\vx\in\mathbb{R}^N$ is the signal of interest with mostly zero entries, and $\vw\in\mathbb{R}^M$ is the measurement noise.

In compressive sensing, we aim to recover the sparse signal $\vx$ given $\{\vy,\vA\}$, with $M\ll N$. In this case, the sensing matrix $\vA$ contains more columns than rows, and there are more than one solutions satisfying the constraint $\|\vy-\vA\vx\|_2^2\leq\epsilon$, where $\epsilon\geq0$ is an upper bound on the measurement noise. This makes the recovery of $\vx$ an ill-posed problem. On the other hand, since the signal of interest itself is sparse, the most straight-forward way to obtain a solution is to pick the one that also shares this ``sparse'' property. Following the well known Occam's razor, we can simply use the $l_0$ norm as the criterion and choose the sparsest (simplest) one:
\begin{align}
P_0^\epsilon(\vx):\quad\min_{\vx}\,\|\vx\|_0\quad\textnormal{subject to }\|\vy-\vA\vx\|_2^2\leq\epsilon\,.
\end{align}
This rather na{\"i}ve attempt is actually backed up by the theoretical work of Cand\`{e}s et al. in \cite{Decode05,RUP06,SRRP06,RIP08}. Under noiseless conditions, it can be shown that the sparsest solution is indeed the true signal when $\vx$ is sufficiently sparse and $\vA$ satisfies the corresponding restricted isometry property \cite{Decode05,RIP08}.

\begin{figure*}[tbp]
\centering
\subfigure{
\label{fig:shannon_entropy_2d_10}
\includegraphics[width=2.25in]{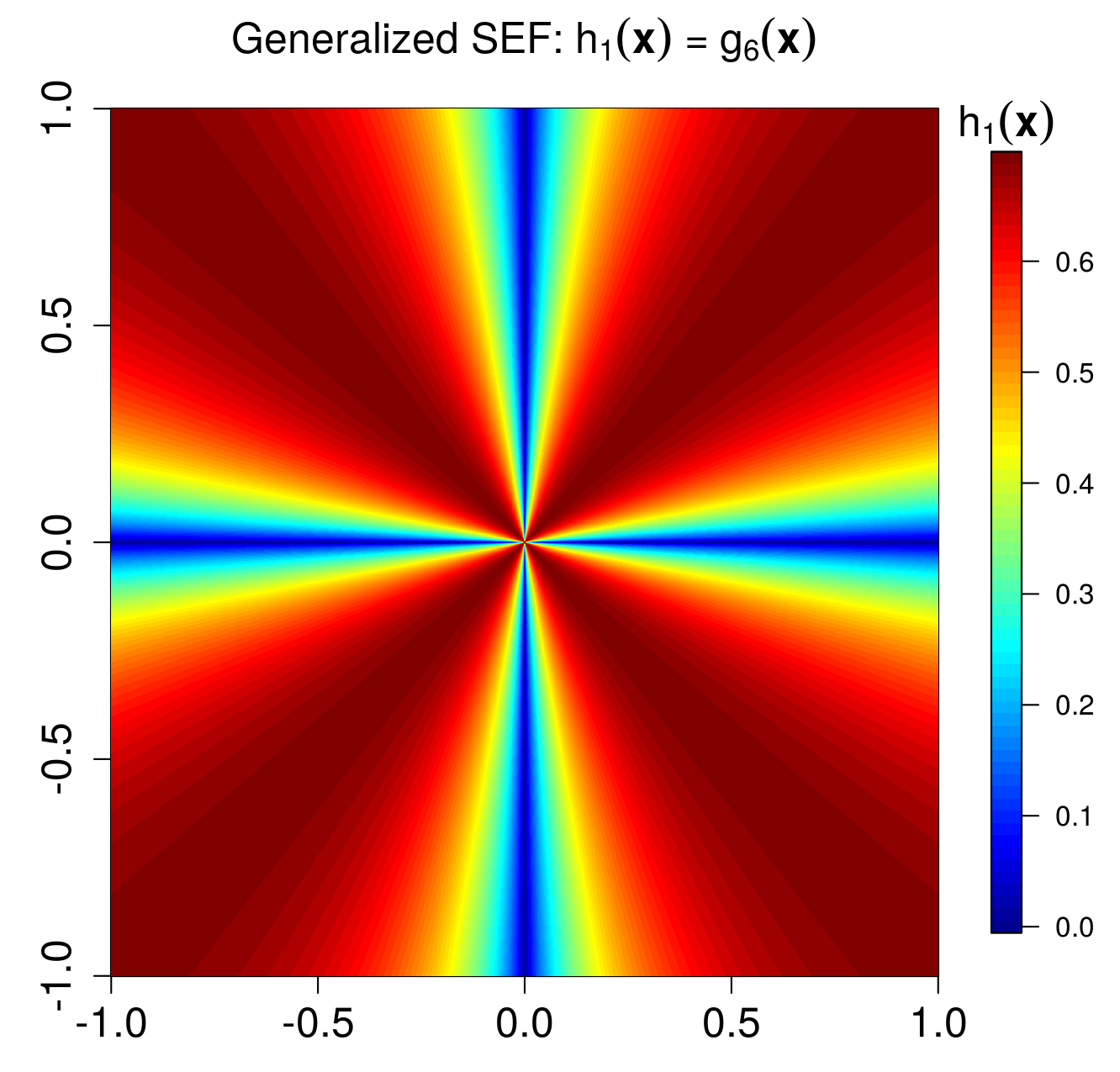}}
\subfigure{
\label{fig:shannon_entropy_2d_20}
\includegraphics[width=2.25in]{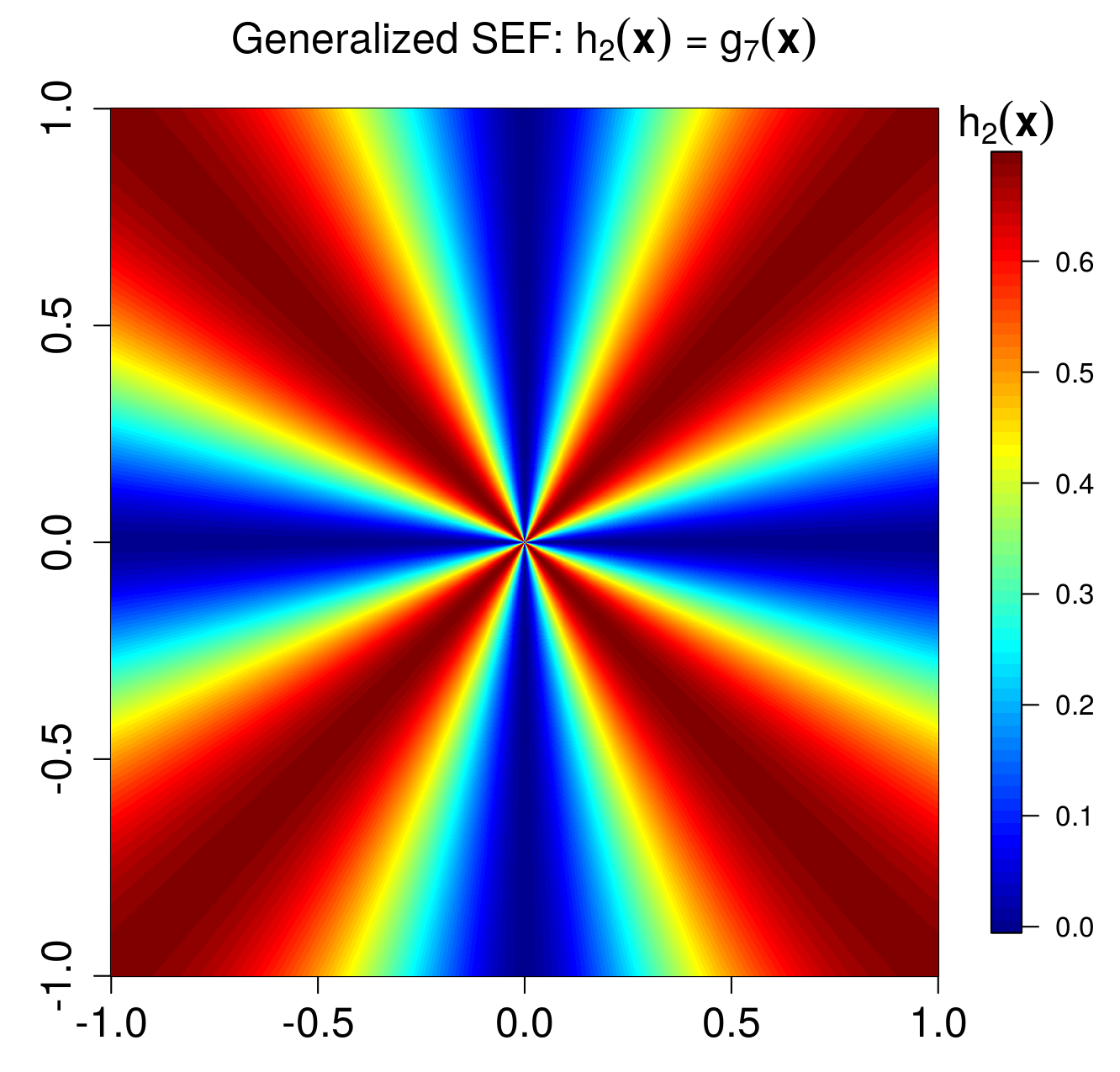}}
\subfigure{
\label{fig:l10_2d}
\includegraphics[width=2.25in]{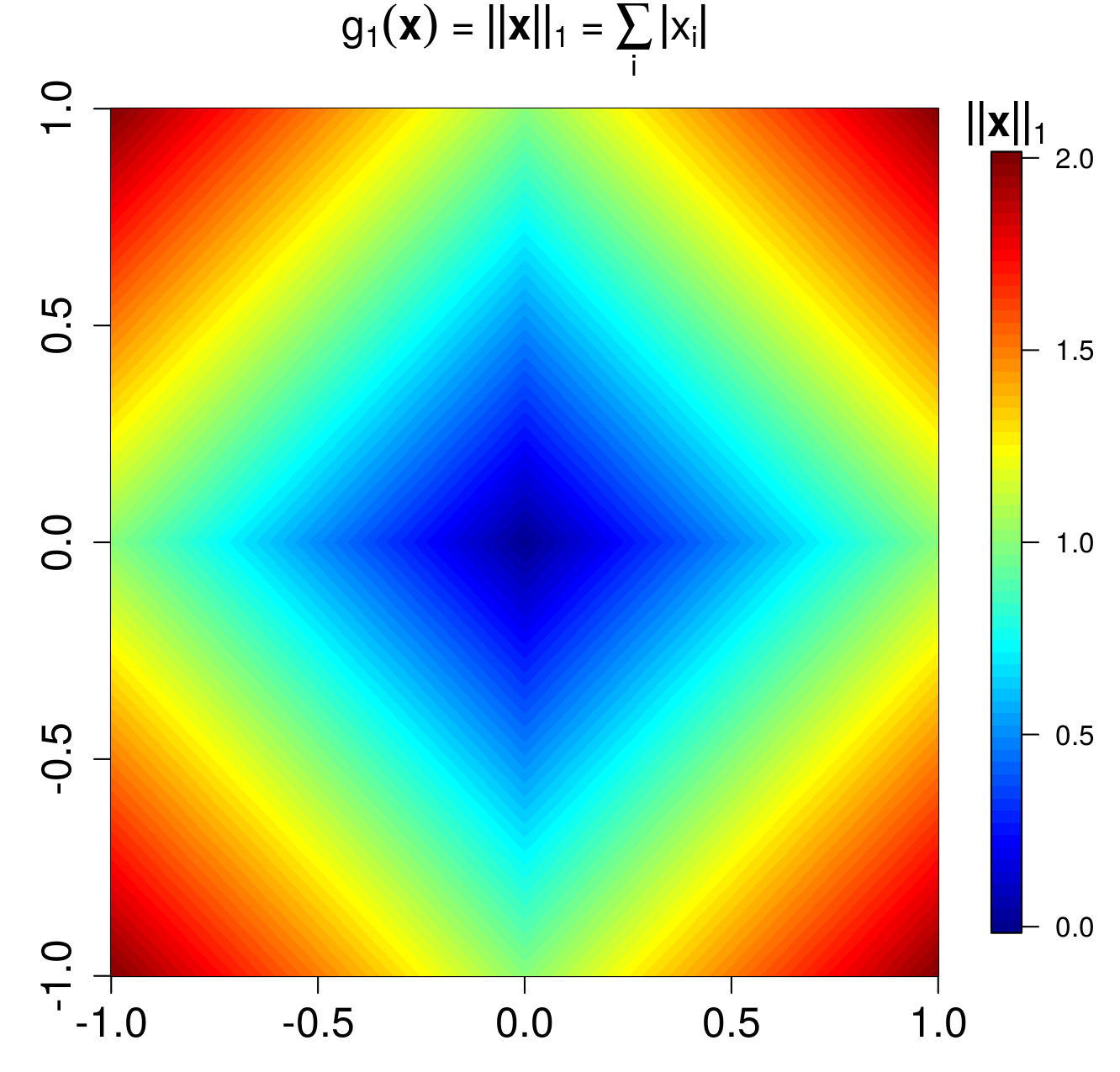}}\\
\vspace{-1mm}
\subfigure{
\label{fig:l05_2d}
\includegraphics[width=2.25in]{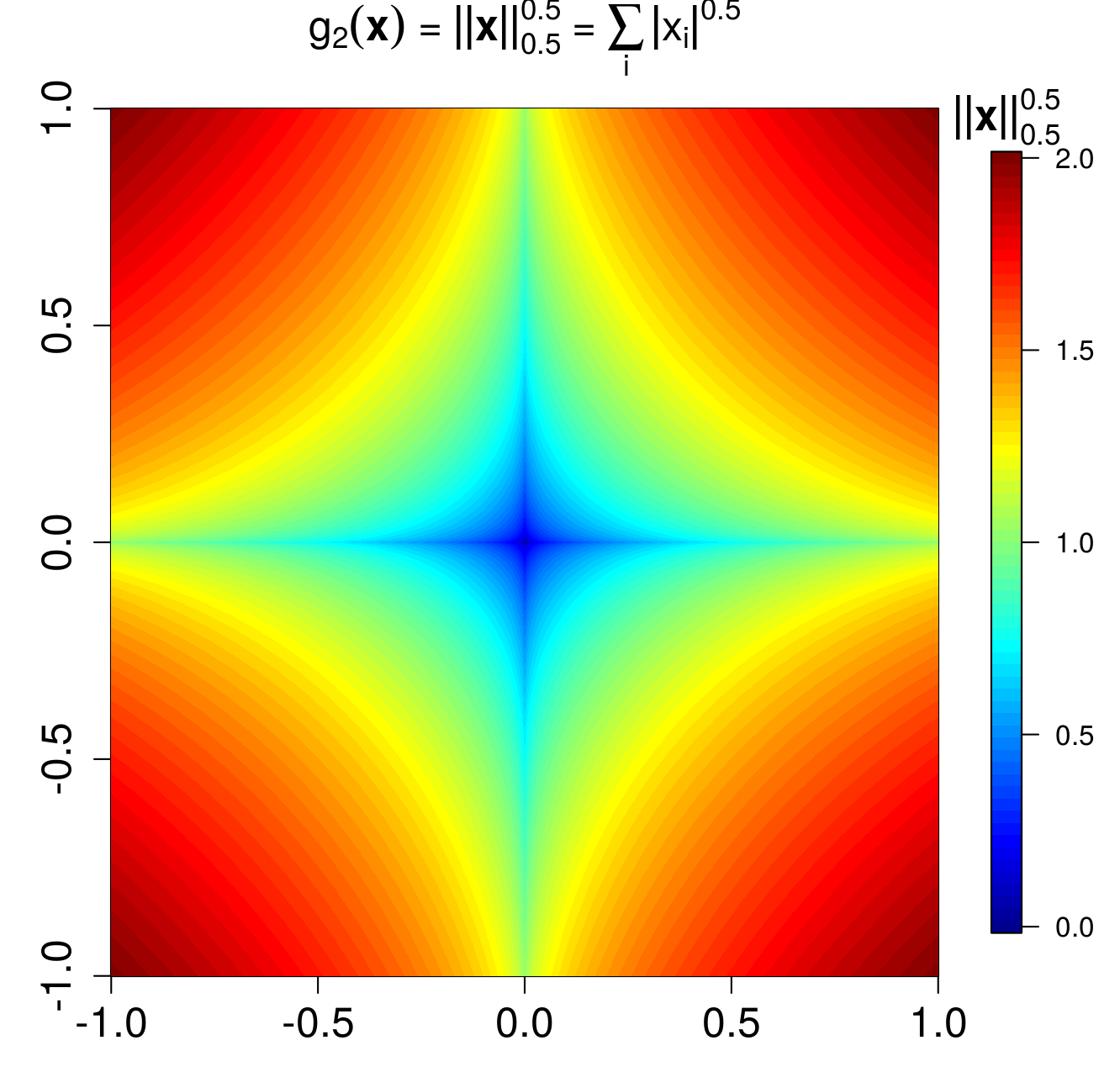}}
\subfigure{
\label{fig:l20_2d}
\includegraphics[width=2.25in]{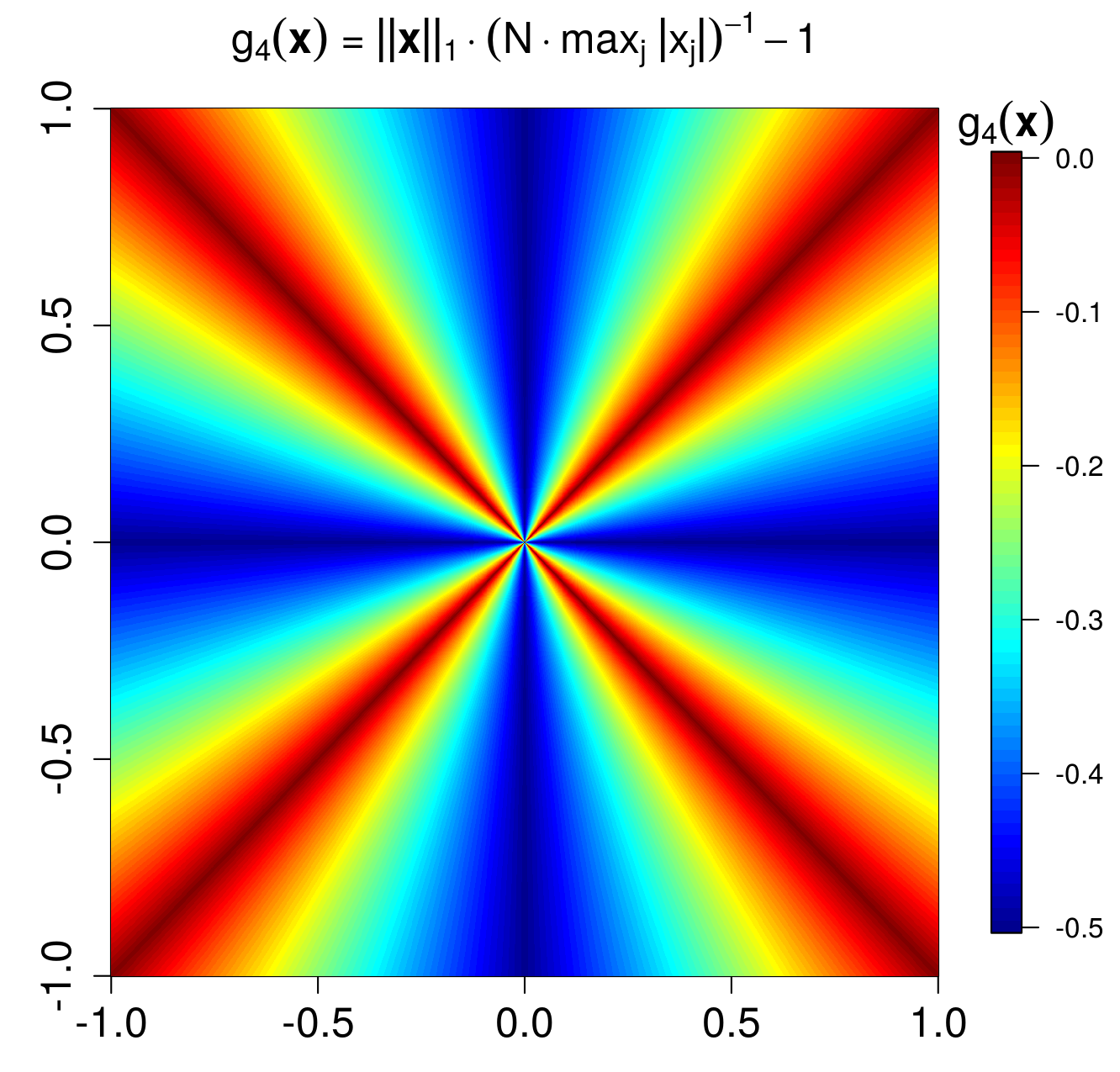}}
\subfigure{
\label{fig:l05_2d}
\includegraphics[width=2.25in]{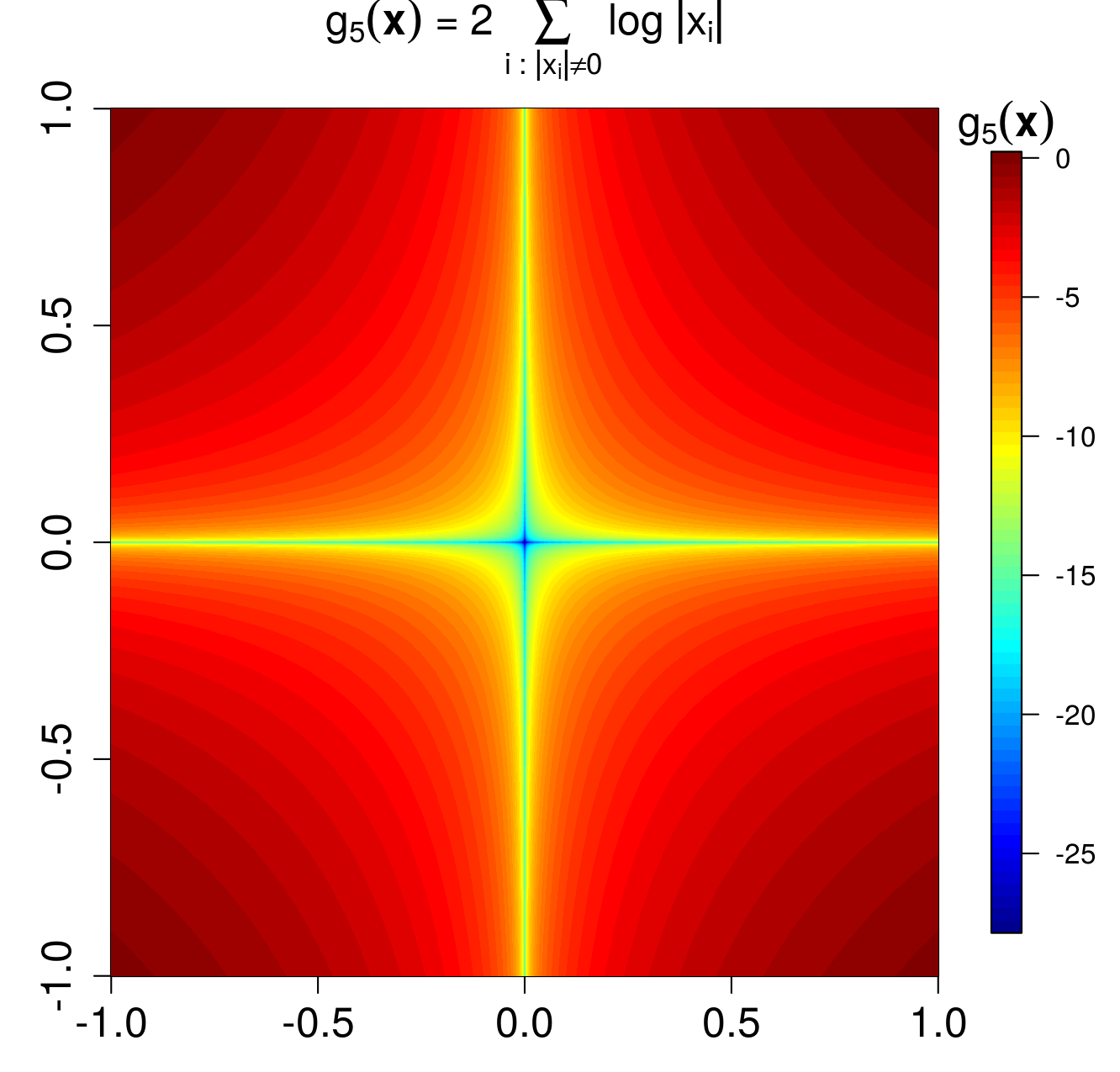}}
\vspace{-2mm}
\caption{The proposed generalized Shannon entropy function (SEF) $h_p(\vx)$ compared with various sparsity-promoting regularization functions introduced in section \ref{sec:sparsity_regularizers} in $\mathbb{R}^2$.}
\vspace{-2mm}
\label{fig:various_functions}
\end{figure*}

Alternatively, we could impose an upper bound $S\in\mathbb{N}^+$ on the sparsity level of $\vx$, and minimize the data fidelity term:
\begin{align}
\label{eq:l0_S_con}
P_0^S(\vx):\quad\min_{\vx}\,\|\vy-\vA\vx\|_2^2\quad\textnormal{subject to }\|\vx\|_0\leq S\,.
\end{align}
Both $P_0^\epsilon(\vx)$ and $P_0^S(\vx)$ are \emph{nonconvex} NP-hard problems in general \cite{SAS95,Davis94}. The $l_0$ norm can also be used as a regularizer \cite{Blumensath2008}, which gives rise to the following unconstrained problem:
\begin{align}
\label{eq:l0_lambda_uncon}
P_0^\lambda(\vx):\quad\min_{\vx}\,\|\vy-\vA\vx\|_2^2+\lambda\|\vx\|_0\,,
\end{align}
where $\lambda>0$ is the regularization parameter. 

In practice, the following two approaches are usually employed to solve the sparse signal recovery problem:
\begin{enumerate}
\item Minimization of the $l_0$ norm $\|\vx\|_0$. 
\begin{itemize}
\item Various pursuit methods directly search for the nonzero entries of $\vx$ incrementally or collectively. 

Let $\widetilde{\vx}$ denote the nonzero entries and $\widetilde{\vA}$ denote the corresponding support set, matching pursuit (MP) \cite{MP95} greedily searches for the sparsest representation of $\vy$ by adding one column of $\vA$ to $\widetilde{\vA}$ at a time until the residue is no larger than $\epsilon$ or $\|\vx\|_0$ reaches $S$. The corresponding coefficient is also added to $\widetilde{\vx}$ one by one during the process. Orthogonal matching pursuit \cite{OMP07,Cai:OMP:2011} is an extension of MP, it updates all the coefficients in $\widetilde{\vx}$ with the orthogonal projection of $\vy$ onto $\widetilde{\vA}$ in every iteration. 

If the sparsity level $S$ is known, the support set $\widetilde{\vA}$ can be further updated collectively during every iteration using CoSaMP \cite{CoSaMP09} or subspace pursuit \cite{SP09}.

\item In an effort to solve \eqref{eq:l0_S_con} and \eqref{eq:l0_lambda_uncon}, Blumensath and Davis \cite{Blumensath2008,Blumensath2009,Blumensath2010} proposed the iterative hard thresholding (IHT) to refine the solution $\vx^{(t+1)}$ iteratively:
\begin{align}
\vx^{(t+1)} = \textnormal{H}\left(\vx^{(t)}+\vA^\textnormal{T}(\vy-\vA\vx^{(t)})\right)\,,
\end{align}
where $\textnormal{H}(\vx)$ is a nonlinear operator defined as follows:
\begin{enumerate}[label=\alph*)]
\item $\textnormal{H}_S(\vx)$ preserves the largest $S$ entries of $\vx$ in magnitude and sets the rest to $0$.
\item $\textnormal{H}_\lambda(\vx)$ preserves the entries whose magnitude is no less than $\lambda$ and sets the rest to $0$.
\end{enumerate}
The convergence of IHT is guaranteed provided that the operator norm $\|\vA\|_2$ is less than $1$ \cite{Blumensath2009}. IHT can be generalized to recover a row-sparse matrix $\vX$ \cite{JointIHT2011}.
\end{itemize}
The various pursuit methods and IHT can be further modified under the oblique pursuits framework \cite{OblPs2013} to work with a wider class of measurement matrices.
\item Relaxation of the $l_0$ norm $\|\vx\|_0$.

The relaxed alternative objective functions should be able to quantify the sparsity level of $\vx$ and easy to optimize \cite{SparseMeasure09}. A detailed introduction is deferred to section \ref{sec:sparsity_regularizers}.
\end{enumerate}

Here we focus on studying the ``relaxation'' approach that solves the following \emph{unconstrained} recovery problem:
\begin{align}
\label{eq:sparsity_regularization}
P_g(\vx):\quad\min_{\vx}\,\|\vy-\vA\vx\|_2^2+\lambda\, g(\vx)\,,
\end{align}

where $\lambda>0$ is the parameter that balances the trade-off between the data fidelity term $\|\vy-\vA\vx\|_2^2$ and the sparsity regularizer $g(\vx)$. The sparse prior is enforced via the regularizer $g(\vx)$. A proper $g(\vx)$ is crucial to the success of the sparse signal recovery task: it should favor sparse solutions, and the corresponding problem $P_g(\vx)$ can be solved efficiently. 

Built upon our earlier work in \cite{EFM16}, in this paper we propose the following \emph{generalized} Shannon entropy function and R{\'e}nyi entropy function as new sparsity-promoting regularizers:
\begin{enumerate}
\item Shannon entropy function:
\begin{align}
\label{eq:shannon_entropy_function}
h_p(\vx)=-\sum_{i=1}^N\frac{|x_i|^p}{\|\vx\|^p_p}\log\frac{|x_i|^p}{\|\vx\|^p_p}\,,
\end{align}
where $p>0$.
\item R{\'e}nyi entropy function:
\begin{align}
\label{eq:renyi_entropy_function}
h_{p,\alpha}(\vx)=\frac{1}{1-\alpha}\log\left(\sum_{i=1}^{N}\left(\frac{|x_i|^p}{\|\vx\|_p^p}\right)^\alpha\right)\,,
\end{align}
where $p>0$, $\alpha>0$ and $\alpha\neq1$.
\end{enumerate}
Both $h_p(\vx)$ and $h_{p,\alpha}(\vx)$ are nonconvex functions, scale-independent, and their local minimums occur at the boundary of each orthant in the Euclidean space $\mathbb{R}^N$, i.e. the axes. They share similar structures in $\mathbb{R}^N$. Take the Shannon entropy function for example, the 2D level plots of $h_p(\vx)$ with $p=\{1,\,2\}$ are shown in Fig. \ref{fig:various_functions}. In this paper we show their advantages over other popular regularizers in promoting sparse solutions with both theoretical analyses and experimental evaluations.

\begin{table*}[tbp]
\caption{Prior entropy-based functionals.}
\label{tab:entropy_functionals}
\centering
\begin{tabular}{cccc}
\toprule
$g_6(\vx)$\cite{UCSD97} &$g_7(\vx)$\cite{Entropy92,WaveletAnalysis94,UCSD97,ScaleBasis99} &$g_8(\vx)$\cite{MES94} &$g_9(\vx)$\cite{UCSD97}  \\ \midrule
$-\sum_i\frac{|x_i|}{\|\vx\|_1}\log\frac{|x_i|}{\|\vx\|_1}$ &$-\sum_i\frac{|x_i|^2}{\|\vx\|_2^2}\log\frac{|x_i|^2}{\|\vx\|_2^2}$ &$-\sum_i|x_i|\log|x_i|$ &$-\sum_i|x_i|^2\log|x_i|^2$\\
$\vx\in\mathbb{R}^N$, $\vx\neq 0$ &$\vx\in\mathbb{R}^N$, $\vx\neq 0$ &$\vx\in\mathbb{R}^N$, $\|\vx\|_1$ is fixed. &$\vx\in\mathbb{R}^N$, $\|\vx\|_2^2$ is fixed.
\\ \bottomrule
\end{tabular}
\end{table*}

\subsection{Prior Work}
\subsubsection{Sparsity-promoting regularizers}
\label{sec:sparsity_regularizers}
Various regularizers have been proposed as the relaxations of the $l_0$ norm. 
\begin{enumerate}[label=\alph*)]
\item Most popular among them are the $l_p$-norm based functions, i.e. the \emph{convex} $l_1$ norm $\|\vx\|_1$ \cite{WaveletAnalysis94,MES94,l1stable06} and the \emph{nonconvex} $p$-th power of the $l_p$ norm $\|\vx\|_p^p$, $0<p<1$\ \cite{ScaleBasis99,Nonconvex_lp07,Lp08,IRFSS10}:
\begin{itemize}
\item $g_1(\vx)=\|\vx\|_1=\sum_i|x_i|$, where $\vx\in\mathbb{R}^N$.
\item $g_2(\vx)=\|\vx\|_p^p=\sum_i|x_i|^p$, where $\vx\in\mathbb{R}^N$, $p\in(0,1)$.
\end{itemize}

Both $\|\vx\|_1$ and $\|\vx\|_p^p$ are additive separable functions, and strict error bounds can be established for the recovered solutions from $l_1$-minimization \cite{Decode05,RUP06,SRRP06,RIP08} and $l_p^p$-minimization problems \cite{Nonconvex_lp07,RIPlp08,Lp08}. Compared to $l_1$-minimization, $l_p^p$-minimization has a tighter error bound and better sparse recovery performances.

\item When $\vx$ lies in the probability simplex, i.e. $\|\vx\|_1=1$ and $x_i\geq 0,\forall i$, minimizing $\|\vx\|_1$ would not work in this case. \cite{NIPS2012_card} proposed a lower bound $g_3(\vx)$ on $\|\vx\|_0$, and showed that the optimization problem can be efficiently solved using convex programming: 
\begin{itemize}
\item $g_3(\vx)=\frac{1}{\|\vx\|_\infty}=\frac{1}{\max_jx_j}\leq\|\vx\|_0$, 

where $\vx>\boldsymbol{0}$, $\|\vx\|_1=1$.
\end{itemize}
\cite{SpaEnt2013} later proposed the following sparsity regularizer\footnote{The sparsity measure proposed in \cite{SpaEnt2013} is actually $-g_{4}(\vx)$. The sign is flipped because we are minimizing the regularizer here.}:
\begin{itemize}
\item $g_{4}(\vx)=\frac{\|\vx\|_1}{N\cdot\|\vx\|_\infty}-1=\frac{\|\vx\|_1}{N\cdot\max_j |x_j|}-1$,

where $\vx\in\mathbb{R}^N$, $\vx\neq\boldsymbol 0$.
\end{itemize}
When $\|\vx\|_1=1$ and $\vx>0$, minimizing $g_{4}(\vx)$ is equivalent to minimizing $g_3(\vx)$. The function $g_3(\vx)$ could be considered to be a modified special case of $g_{4}(\vx)$. For readability purposes, we shall refer to $g_4(\vx)$ as the ``$\textnormal{L}_1/\textnormal{L}_\infty$'' function in this paper.

Although the $\textnormal{L}_1/\textnormal{L}_\infty$ function and the proposed $h_p(\vx)$, $h_{p,\alpha}(\vx)$ are all scale-independent (see Fig. \ref{fig:various_functions} for a 2D example), they differ in the following aspects:
\begin{itemize}[label={--}]
\item The proposed entropy functions are continuously differentiable w.r.t. the modulus $|\vx|$, while the $\textnormal{L}_1/\textnormal{L}_\infty$ function is not.
\item  The proposed entropy functions have adaptive energy-promoting behaviors, while the $\textnormal{L}_1/\textnormal{L}_\infty$ function has a rigid energy-promoting behavior. The detailed analysis is presented in section \ref{sec:energy_promoting} and Appendix \ref{app:energy_promoting_other}.
\end{itemize}

\item Another popular class of regularizers is constructed from the logarithm function $\log(\cdot)$ \cite{WaveletAnalysis94,Figueiredo:2003,Rao:Subset:2003,ScaleBasis99}. The following ``logarithm of energy'' function was proposed in \cite{WaveletAnalysis94} as a measure of sparsity:
\begin{itemize}
\item $g_5(\vx)=\sum_{i:|x_i|\neq 0}\log|x_i|^2=2\sum_{i:|x_i|\neq 0}\log|x_i|$, where $\vx\in\mathbb{R}^N$, $\vx\neq\boldsymbol 0$.
\end{itemize}
Cand\`{e}s et al. \cite{RWL108} later proposed the reweighted $l_1$-minimization algorithm as a way to enhance the sparsity, which is essentially the iterative minimization of the first-order approximation of $g_5(\vx)$: $\sum_i\log(|x_i|+\epsilon)$, where $\epsilon>0$ is some small positive constant.

\item As listed in Table \ref{tab:entropy_functionals}, entropy-based functionals have also been widely used to promote sparsity in $\vx$.  Specifically, $g_6(\vx)$ and $g_7(\vx)$ are special cases of the proposed generalized Shannon entropy function $h_p(\vx)$ in \eqref{eq:shannon_entropy_function} when $p=1$ and $2$. There is some imprecision in \cite{UCSD97}'s analysis on the local minimums of $g_7(\vx)$, which were believed to ``occur just shy of the boundaries defined by the coordinate axes''. In section \ref{subsec:sparse_analysis}, we can prove that the local minimums actually occur exactly on the coordinate axes. $g_8(\vx)$ only promotes sparsity if the signal $\vx$ has fixed $l_1$ norm, while $g_9(\vx)$ works for signals with fixed $l_2$ norm.



\item The functions $g_6(\vx)$, $\cdots$, $g_9(\vx)$ are all nonconvex. \cite{Kose2014} later proposed the following convex entropy function as an approximation to the $l_1$-norm $\|\vx\|_1$:
\begin{itemize}
\item $g_{10}(\vx)=\sum_i\left[(|x_i|+e^{-1})\log(|x_i|+e^{-1}) \,+\, e^{-1}\right]$,\\ where $\vx\in\mathbb{R}^N$.
\end{itemize}
This way $g_{10}(\vx)$ maintains the strictly convex property of $\|\vx\|_1$ and is continuously differentiable in $\mathbb{R}^N$. However, we should note that $g_{10}(\vx)$ only produces concentrated but not truly sparse solutions.
\end{enumerate}

\subsubsection{$l_1$-norm minimization algorithms}
When $g(\vx)=\|\vx\|_1$, the \emph{convex} $l_1$-norm minimization problem can be efficiently solved using various readily available algorithms such as GPSR \cite{GPSR07}, ISTA \cite{Wavelet98,ISTA03,ISTA04}, TwIST\cite{TwIST07}, FISTA \cite{FISTA09,Beck2009FastGA}, NESTA \cite{Nesterov83,NESTA11}, ADM\cite{ADMM11}, etc. Detailed reviews of popular $l_1$-norm minimization algorithms can be found in \cite{Yang2010ARO,L1review13}.

In this paper the optimization problem \eqref{eq:sparsity_regularization} is recasted into a series of reweighted $l_1$ problems \cite{RWL108,IRFSS10}. We choose to adapt the FISTA to solve \eqref{eq:sparsity_regularization} for its easy implementation and fast convergence.

\subsubsection{Uniqueness of the solution}
\label{sec:uniq_sol}
The restricted isometry property (RIP) of the $M\times N$ sensing matrix $\vA$ was introduced in \cite{Decode05} to provide guarantees for exact recovery. Let $\delta_S$ be the smallest number such that:
\begin{align}
(1-\delta_S)\|\vx\|_2^2\leq\|\vA\vx\|_2^2\leq(1+\delta_S)\|\vx\|_2^2\,.
\end{align}
holds for all vectors $\vx$ with $\|\vx\|_0\leq S$, then the matrix $\vA$ is said to satisfy RIP with $S$-restricted constant $\delta_S$. In the noiseless case \cite{RIP08}, when $\delta_{2S}<\sqrt{2}-1$, minimizing the convex relaxation $\|\vx\|_1$ is equivalent to minimizing the nonconvex $\|\vx\|_0$, and has a unique $S$-sparse solution. In the noisy case \cite{RIP08}, a strict upper bound on the $l_2$ error $\|\hat{\vx}-\vx\|$ of the recovered $\|\vx\|_1$-minimization solution $\hat{\vx}$ can be established. 

It was shown in \cite{RIP08} that the RIP can be met with high probability for Gaussian random matrix with i.i.d entries and $M>CS\log(N/S)$ for some constant $C$. Chartrand et al. \cite{Nonconvex_lp07,RIPlp08} later gave the uniqueness conditions for $\|\vx\|_p^p$-minimization in terms of the RIP of $\vA$. Both $\|\vx\|_1$ and $\|\vx\|_p^p$ are additive separable functions in that they can be written as a sum of identical functions involving individual entries of $\vx$. However, the proposed entropy functions are non-separable, and different approaches are needed to find the properties of $\vA$ that could guarantee the uniqueness of the solution. 
 
\subsection{Main Contributions}
Compared to previously adapted entropy functions \cite{Entropy92,WaveletAnalysis94,UCSD97,ScaleBasis99} where only $l_1$ and $l_2^2$ normalized $\vx$ were studied, the proposed Shannon entropy function $h_p(\vx)$ and R\'{e}nyi entropy function $h_{p,\alpha}(\vx)$ in (\ref{eq:shannon_entropy_function}, \ref{eq:renyi_entropy_function}) are more \emph{general}. They are defined with respect to the probability distribution in (\ref{eq:prob_construct}) where $p>0$ can be any positive number. This gives us more freedom in constructing the proper entropy functions for the sparse signal recovery task. As evident from the experiments on both simulated and real data, the ability to choose suitable $p$ values enables us to fully exploit the sparsity-promoting power of the generalized entropy functions and to achieve better performances over other state-of-the-art regularized approaches.

Previous works \cite{UCSD97,ScaleBasis99} focused on the study of the Schur concavity of the generalized entropy functions with respect to $\vx$ where $p=1,\,2$. They believed that the Shannon entropy function produced truly sparse solutions only when $p=1$. Here we show that it's actually the (Schur) concavity with respect to the \emph{distribution} $\mathscr{P}(\cdot)$ that really matters. The Shannon entropy function with $p=1$ is not the only case where truly sparse solutions can be obtained. In fact, $\forall\, p>0$ and $0<\alpha\neq 1$, we can prove in section \ref{subsec:sparse_analysis} that the local minimums of the generalized entropy functions only occur at the boundaries of the orthants in $\mathbb{R}^N$. Hence minimizing $h_p(\vx)$ or $h_{p,\alpha}(\vx)$ in said orthant $\mathbb{O}$ will lead us to the solutions on its boundary, i.e. sparser solutions.

Additionally, minimizing the generalized entropy functions adaptively promotes multiple high-energy coefficients while suppressing the rest low-energy coefficients. Combining the proximal approximation \cite{Martinet1970, Proximal13} of the data fidelity term and the first order approximations of the generalized entropy functions, we can recast the nonconvex entropy functions minimization problems into a series of simple reweighted $l_1$ problems, and solve them efficiently via the accelerated inexact proximal gradient method \cite{FISTA09,Beck2009FastGA,APG2015}. 

\section{Sparsity-Regularization Entropy Functions}
\subsection{Entropy in Information Theory}
Here we first introduce the entropy concepts in information theory \cite{Cover06,renyi60entropy}. Both the Shannon entropy and R\'{e}nyi entropy are defined with respect to the probability distribution $\boldsymbol{\mathscr{P}}(\mathcal{V})$ of some random variable $\mathcal{V}$. We give the definitions in terms of discrete probability distribution\footnote{For continuous distributions, the sum $\sum$ in (\ref{eq:shannon_entropy},\ref{eq:renyi_entropy}) should be replaced with integration $\int$. } as follows:
\begin{itemize}
\item Shannon entropy\footnote{The ``$\log$'' in this paper is by default natural logarithm, i.e. base $e$}:
\begin{align}
\label{eq:shannon_entropy}
\mathcal{H}(\mathcal{V})=-\textstyle\sum_{i=1}^{|\mathcal{V}|}\mathscr{P}(v_i)\log \mathscr{P}(v_i)\,.
\end{align}
$\mathcal{H}(\mathcal{V})$ is strictly concave with respect to the probability distribution $\boldsymbol{\mathscr{P}}(\mathcal{V})=\{\mathscr{P}(v_1),\cdots,\mathscr{P}(v_{|\mathcal{V}|})\}$.

\item R\'{e}nyi entropy:
\begin{align}
\label{eq:renyi_entropy}
\mathcal{H}_\alpha(\mathcal{V})=\frac{1}{1-\alpha}\log\left(\textstyle\sum_{i=1}^{|\mathcal{V}|}\mathscr{P}(v_i)^\alpha\right)\,,
\end{align}
where $\alpha\geq0$ and $\alpha\neq1$. When $\alpha\in(0,1)$, $\mathcal{H}_\alpha(\mathcal{V})$ is strictly concave with respect to $\boldsymbol{\mathscr{P}}(\mathcal{V})$ \cite{Renyi78}; when $\alpha\in(1,\infty)$, $\mathcal{H}_\alpha(\mathcal{V})$ is strictly Schur concave with respect to $\boldsymbol{\mathscr{P}}(\mathcal{V})$ \cite{Bosyk2016}.
\end{itemize}
We should make it clear that Shannon entropy $\mathcal{H}(\mathcal{V})$ is \emph{not} a special case of the R\'{e}nyi entropy, but the limiting value of the R\'{e}nyi entropy $\mathcal{H}_\alpha(\mathcal{V})$ as $\alpha\rightarrow 1$ \cite{renyi_tina04}. Hence we will discuss them separately in this paper.

\subsection{Generalized Entropy Functions of the Sparse Signal}
Entropy measures the uncertainty about the random variable $\mathcal{V}$ with cardinality $|\mathcal{V}|=N$. The lower the entropy is, the more predictable the variable $\mathcal{V}$ will be. This corresponds to a skewed distribution $\boldsymbol{\mathscr{P}}(\mathcal{V})$. The idea of a skewed distribution could translate naturally to a sparse probability vector $\boldsymbol{\mathscr{P}}_{\mathcal{V}}=\left[\mathscr{P}(v_1),\cdots,\mathscr{P}(v_N)\right]^\textnormal{T}$ in the sense that only a few probability entries of $\boldsymbol{\mathscr{P}}_{\mathcal{V}}\in\mathbb{R}^N$ are significant. In other words, the entropy can be used as a measure of how sparse the probability vector $\boldsymbol{\mathscr{P}}_{\mathcal{V}}$ is. This observation motivates us to adapt the concept of entropy as a sparsity-measure for the general signal $\vx$, and use it as a regularizer in the sparse signal recovery task.

As mentioned before, the entropy is defined with respect to some probability distribution $\boldsymbol{\mathscr{P}}(\cdot)$. Here we can construct the following discrete probability distribution out of the signal of interest $\vx\in\mathbb{R}^N$:
\begin{align}
\label{eq:prob_construct}
\vx\rightarrow\left[\frac{|x_1|^p}{\|\vx\|_p^p}, \frac{|x_2|^p}{\|\vx\|_p^p},\cdots,\frac{|x_N|^p}{\|\vx\|_p^p}\right]\,,
\end{align}
where $p>0$. The adaptation from the classical entropy to the generalized entropy function is then pretty straightforward: we simply plug \eqref{eq:prob_construct} into \eqref{eq:shannon_entropy} and \eqref{eq:renyi_entropy} to obtain the generalized Shannon entropy function \eqref{eq:shannon_entropy_function} and the generalized R{\'e}nyi entropy function \eqref{eq:renyi_entropy_function}.

Here we should \emph{not} confuse the ``entropy function'' of $\vx$ with the ``entropy'' of $\vx$. Take the Shannon entropy of $\vx$ for example: $\mathcal{H}(\vx)=-\int_{\vx} \mathscr{P}(\vx)\log \mathscr{P}(\vx)\,d\vx$. $\mathcal{H}(\vx)$ measures the uncertainty of the random variable $\vx$, it is different from $h_p(\vx)$ in \eqref{eq:shannon_entropy_function} that measures the sparsity of $\vx$.

In order to promote sparsity in the recovered solutions, we are going to minimize the generalized entropy functions. The sparse signal recovery problems in the form of (\ref{eq:sparsity_regularization}) using the Shannon entropy function (SEF) $h_p(\vx)$ and the R{\'e}nyi entropy function (REF) $h_{p,\alpha}(\vx)$ as regularizers then become:
\begin{align}
\label{eq:sefm_ssr}
P_{h_p}(\vx):& \quad \min_{\vx} \, \|\vy-\vA\vx\|_2^2+\lambda\,\, h_p(\vx)\\
\label{eq:refm_ssr}
P_{h_{p,\alpha}}(\vx):& \quad \min_{\vx} \, \|\vy-\vA\vx\|_2^2+\lambda\,\, h_{p,\alpha}(\vx)\,.
\end{align}

\subsection{Sparsity-promoting Analysis}
\label{subsec:sparse_analysis}
We next show that $h_p(\vx)$ and $h_{p,\alpha}(\vx)$ can be used as sparsity regularizers in the following sense: minimizing them in an orthant $\mathbb{O}$ of the Euclidean space $\mathbb{R}^N$ leads to solutions on the boundary of said orthant, i.e. sparser solutions.


\begin{noiseless_r}
In this case we are minimizing $h_p(\vx)$ or $h_{p,\alpha}(\vx)$ subject to the constraint $\vy=\vA\vx$, $\vy\neq 0$. We first show that there is a one to one mapping in each orthant between $\vx=\left[x_1,\cdots,x_N\right]^\textrm{T}$ and $\ddot{\vx}=\left[\ddot{x}_1,\cdots,\ddot{x}_N\right]^\textrm{T}$, where $\ddot{x}_i=\textrm{sign}(x_i)\cdot\frac{|x_i|^p}{\|\vx\|^p_p}$. This will be done in two steps: Proposition \ref{proposition:one_to_one_1} and Proposition \ref{proposition:one_to_one_2}.

\begin{proposition}
\label{proposition:one_to_one_1}
If $\vx$ is the solution to $\vy=\vA\vx$, $\vy\neq0$, then there is a one to one mapping in each orthant between $\vx$ and $\tilde{\vx}=\frac{\vx}{\|\vx\|_p}$.
\end{proposition}
\begin{proof}
We need to prove $\vx\longleftrightarrow\tilde{\vx}$:
\begin{itemize}
\item It is easy to verify that $\vx\rightarrow\tilde{\vx}$. 

\item Suppose that $\tilde{\vx}$ can be mapped to two solutions of $\vy=\vA\vx$: $\vx_{(1)},\vx_{(2)}$ in the same orthant. We then have:
\begin{align}
\label{eq:map_same}
\frac{\vx_{(1)}}{\|\vx_{(1)}\|_p}&=\tilde{\vx}=\frac{\vx_{(2)}}{\|\vx_{(2)}\|_p}\\
\frac{\vy}{\|\vx_{(1)}\|_p}=\frac{\vA\vx_{(1)}}{\|\vx_{(1)}\|_p}&=\vA\tilde{\vx}=\frac{\vA\vx_{(2)}}{\|\vx_{(2)}\|_p}=\frac{\vy}{\|\vx_{(2)}\|_p}\,,
\end{align}
which tells us $\frac{\vy}{\|\vx_{(1)}\|_p}=\frac{\vy}{\|\vx_{(2)}\|_p}$. Since $\vy\neq0$, we have $\|\vx_{(1)}\|_p=\|\vx_{(2)}\|_p$. Using (\ref{eq:map_same}), we get $\vx_{(1)}=\vx_{(2)}$. Hence $\vx\leftarrow\tilde{\vx}$.
\end{itemize}
\end{proof}

\begin{figure}[tbp]
\centering
\includegraphics[height=2.1in]{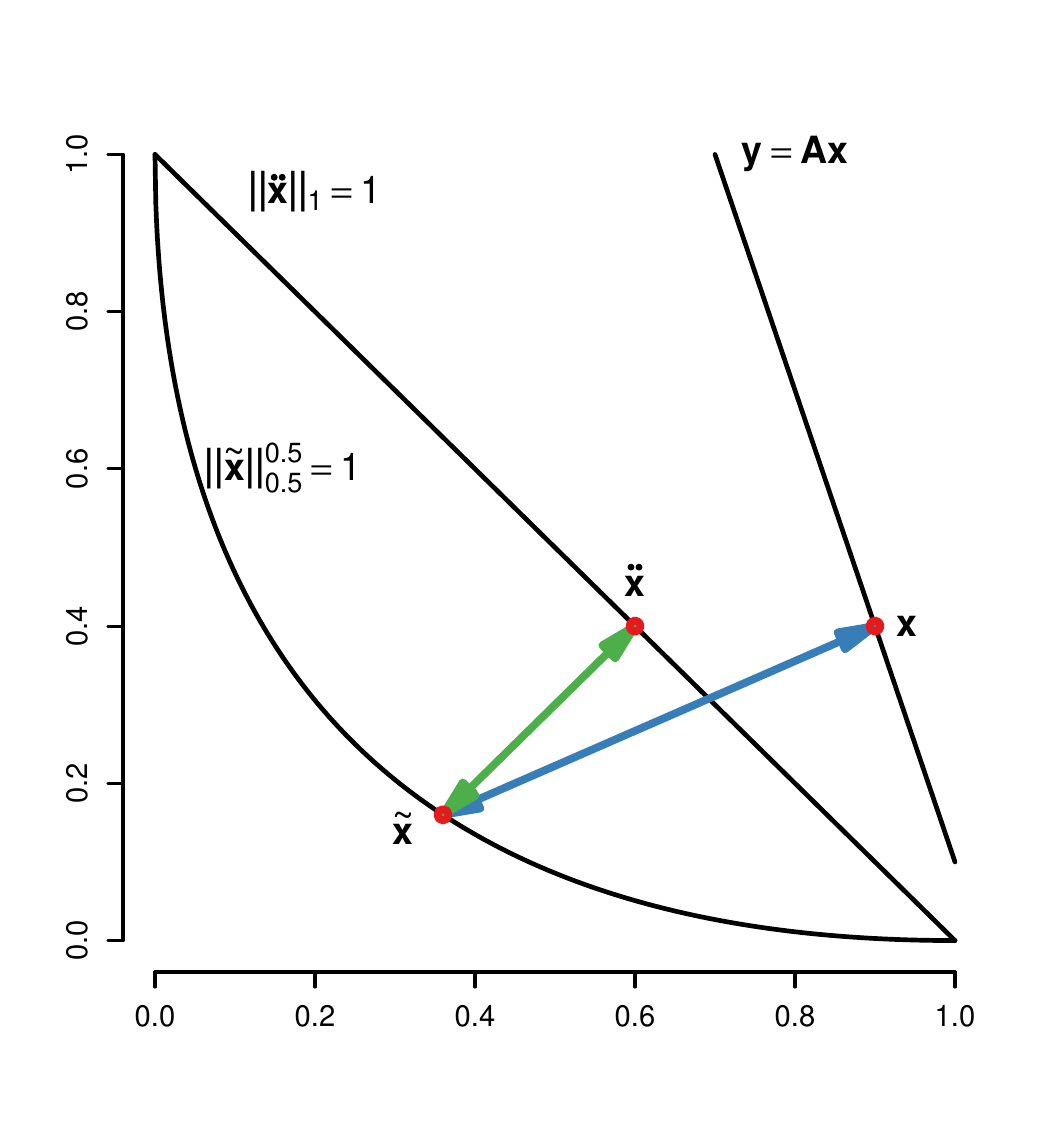}
\caption{The one-to-one mapping in the noiseless case: $\vx\longleftrightarrow\tilde{\vx}\longleftrightarrow\ddot{\vx}$ when $p=0.5$.}
\label{fig:mapping_noiseless}
\vspace{-3mm}
\end{figure}

\begin{proposition}
\label{proposition:one_to_one_2}
\vspace{-2mm}
There is a one to one mapping in each orthant between $\tilde{\vx}$ and $\ddot{\vx}$.
\end{proposition}
\begin{proof}
We need to prove $\tilde{\vx}\longleftrightarrow\ddot{\vx}$:
\begin{itemize}
\item We can rewrite $\ddot{\vx}$ in terms of $\tilde{\vx}$: $\ddot{\vx}=\textrm{sign}(\tilde{\vx})\circ|\tilde{\vx}|^p$, where $\textrm{sign}(\tilde{\vx})$ is a vector of the signs of every entry in $\tilde{\vx}$. Hence $\ddot{\vx}\leftarrow\tilde{\vx}$. 

\item Suppose that $\ddot{\vx}$ can be mapped to two points $\tilde{\vx}_{(1)}, \tilde{\vx}_{(2)}$ in the same orthant. We then have:
\begin{align}
\textrm{sign}(\tilde{\vx}_{(1)})\circ|\tilde{\vx}_{(1)}|^p=\ddot{\vx}=\textrm{sign}(\tilde{\vx}_{(2)})\circ|\tilde{\vx}_{(2)}|^p\,,
\end{align}
where ``$\circ$'' is the element-wise Hadamard product. Since $\textrm{sign}(\tilde{\vx}_{(1)})=\textrm{sign}(\tilde{\vx}_{(2)})$, we can get $|\tilde{\vx}_{(1)}|=|\tilde{\vx}_{(2)}|$.
\begin{align}
\tilde{\vx}_{(1)}=\textrm{sign}(\tilde{\vx}_{(1)})\circ|\tilde{\vx}_{(1)}|=\textrm{sign}(\tilde{\vx}_{(2)})\circ|\tilde{\vx}_{(2)}|=\tilde{\vx}_{(2)}\,.
\end{align}
Hence $\tilde{\vx}_{(1)}=\tilde{\vx}_{(2)}$, and $\ddot{\vx}\rightarrow\tilde{\vx}$.
\end{itemize}
\end{proof}

\vspace{-2mm}
Combining Proposition \ref{proposition:one_to_one_1} and Proposition \ref{proposition:one_to_one_2}, we have $\vx\longleftrightarrow\ddot{\vx}$, as shown in Fig. \ref{fig:mapping_noiseless}. 

Let $\mathcal{X}=\left\{\vx_1,\vx_2,\cdots\right\}$ be the solutions of $\vy=\vA\vx$, $\vy\neq 0$ in one of the orthants $\mathbb{O}$. Specifically, $\mathcal{X}=\mathcal{X}_1\cup\mathcal{X}_2$ and $\mathcal{X}_1\cap\mathcal{X}_2=\emptyset$, where $\mathcal{X}_1$ contains solutions on the boundary of $\mathbb{O}$ and $\mathcal{X}_2$ contains the rest solutions that are not on the boundary. The solution $\vx$ is then mapped to $\ddot{\vx}$ one by one, producing the corresponding sets $\ddot{\mathcal{X}}_1, \ddot{\mathcal{X}}_2$ on the simplex $\|\ddot{x}\|_1=1$. We can verify that the solutions in $\mathcal{X}_1$ are always \emph{sparser} than those in $\mathcal{X}_2$. Additionally, we have the following Proposition \ref{proposition:sparser_solutions}:
\begin{proposition}
\label{proposition:sparser_solutions}
In the noiseless case where $\vy=\vA\vx$, $\vy\neq 0$, for every solution $\vx\in\mathcal{X}_2$, there is a solution $\vx^*\in\mathcal{X}_1$ on the boundary of the orthant $\mathbb{O}$ such that $h_p(\vx^*)<h_p(\vx)$ and $h_{p,\alpha}(\vx^*)<h_{p,\alpha}(\vx)$.
\end{proposition}
\begin{proof}
By definition we have:
\begin{align}
&h_p(\vx)=g(\ddot{\vx})=-\textstyle\sum_{i=1}^N|\ddot{x}_i|\log|\ddot{x}_i|\\
&h_{p,\alpha}(\vx)=g_{\alpha}(\ddot{\vx})=\frac{1}{1-\alpha}\log\left(\textstyle\sum_{i=1}^N|\ddot{x}_i|^\alpha\right)
\,.
\end{align} 

For the SEF $h_p(\vx)$, we first study the local minimums on the simplex $\|\ddot{\vx}\|_1=1$ in the orthant $\mathbb{O}$. $g(\ddot{\vx})$ is \emph{strictly concave} with respect to $\ddot{\vx}$, and the local minimums of $g(\ddot{\vx})$ are on the boundary of $\mathbb{O}$. Hence for every $\ddot{\vx}\in\ddot{\mathcal{X}}_2$, there is a $\ddot{\vx}^*\in\ddot{\mathcal{X}}_1$ such that $g(\ddot{\vx}^*)<g(\ddot{\vx})$. 

For the REF $h_{p,\alpha}(\vx)$, when $\alpha\in(0,1)$, $g_\alpha(\ddot{\vx})$ is \emph{strictly concave} with respect to $\ddot{\vx}$, the local minimums of $g_{\alpha}(\ddot{\vx})$ are on the boundary of the orthant $\mathbb{O}$. When $\alpha\in(1,\infty)$, $g_\alpha(\ddot{\vx})$ is \emph{strictly Schur concave} \cite{Bosyk2016}, since the boundary of the orthant $\mathbb{O}$ majorizes the $\ddot{\vx}$ inside $\mathbb{O}$. The local minimums of $g_{\alpha}(\ddot{\vx})$ are also on the boundary of $\mathbb{O}$. Hence for every $\ddot{\vx}\in\ddot{\mathcal{X}}_2$, there also exists a $\ddot{\vx}^*\in\ddot{\mathcal{X}}_1$ such that $g_\alpha(\ddot{\vx}^*)<g_\alpha(\ddot{\vx})$ for $\alpha\in (0,1)\cup(1,\infty)$. 

Proposition \ref{proposition:one_to_one_1} and \ref{proposition:one_to_one_2} guarantee a one to one mapping in $\mathbb{O}$ between $\vx$ and $\ddot{\vx}$: $\vx\longleftrightarrow\ddot{\vx}$. Since $h_p(\vx)=g(\ddot{\vx})$ and $h_{p,\alpha}(\vx)=g_{\alpha}(\ddot{\vx})$, for every $\vx\in\mathcal{X}_2$, there is an $\vx^*\in\mathcal{X}_1$ such that $h_p(\vx^*)<h_p(\vx)$ and $h_{p,\alpha}(\vx^*)<h_{p,\alpha}(\vx)$.
\end{proof}

From Proposition \ref{proposition:sparser_solutions} we can see that minimizing $h_p(\vx)$ or $h_{p,\alpha}(\vx)$ in the orthant $\mathbb{O}$ leads us to sparser solutions in $\mathcal{X}_1$.
\end{noiseless_r}

\begin{noisy_r}
We can show similarly that minimizing $h_p(\vx)$ or $h_{p,\alpha}(\vx)$ subject to the constraint $\|\vy-\vA\vx\|_2^2\leq\epsilon$ in an orthant $\mathbb{O}$ of the Euclidean space $\in\mathbb{R}^N$ also produces sparse solutions. First, we have the following Proposition \ref{proposition:one_to_one_1_noisy}:
\begin{proposition}
\label{proposition:one_to_one_1_noisy}
Let $\mathcal{X}^\epsilon=\left\{\vx_1,\vx_2,\cdots\right\}$ are the nonzero solutions satisfying the constraint $\|\vy-\vA\vx\|_2^2\leq\epsilon, \vy\neq0$ such that: $\forall \vx_i\neq\vx_j$, $\vx_i=\tau\vx_j$ for some $\tau>0$. Pick any $\vx_i\in\mathcal{X}^\epsilon$, there is a one to one mapping in each orthant between the set $\mathcal{X}^\epsilon$ and $\tilde{\vx}_i=\frac{\vx_i}{\|\vx_i\|_p}$.
\end{proposition}
\begin{proof}
We need to prove $\mathcal{X}^\epsilon\longleftrightarrow\tilde{\vx}_i$:
\begin{itemize}
\item $\forall \vx_j\in\mathcal{X}^\epsilon\backslash\vx_i$, we have $\tilde{\vx}_i=\frac{\vx_i}{\|\vx_i\|_p}=\frac{\tau\vx_j}{\|\tau\vx_j\|_p} = \frac{\vx_j}{\|\vx_j\|_p}=\tilde{\vx}_j$. We can verify that $\vx_j\rightarrow\tilde{\vx}_j=\tilde{\vx}_i$. Hence $\mathcal{X}^\epsilon\rightarrow\tilde{\vx}_i$.
\item Suppose that there are two sets $\mathcal{X}^\epsilon_{(1)},\mathcal{X}^\epsilon_{(2)}$ in the same orthant being mapped to the same $\tilde{\vx}_i$. Let $\vx_1\in\mathcal{X}^\epsilon_{(1)}$ and $\vx_2\in\mathcal{X}^\epsilon_{(2)}$, we have:
\begin{align}
\frac{\vx_1}{\|\vx_1\|_p}=\tilde{\vx}_1=\tilde{\vx}_i=\tilde{\vx}_2=\frac{\vx_2}{\|\vx_2\|_p}\,.
\end{align}
We then have $\vx_1=\frac{\|\vx_1\|_p}{\|\vx_2\|_p}\vx_2$, which means that $\vx_1,\vx_2$ belongs to the same set, i.e. $\mathcal{X}^\epsilon_{(1)}=\mathcal{X}^\epsilon_{(2)}$. Hence $\mathcal{X}\leftarrow\tilde{\vx}_i$.
\end{itemize}
\end{proof}

We can see that the solutions in the same $\mathcal{X}^\epsilon$ all have the same entropy function value. Combining Proposition \ref{proposition:one_to_one_1_noisy} and Proposition \ref{proposition:one_to_one_2}, we have $\mathcal{X}^\epsilon\longleftrightarrow\ddot{\vx}_i$, as illustrated in Fig. \ref{fig:mapping_noisy}. 

\begin{figure}[tbp]
\centering
\includegraphics[height=2.1in]{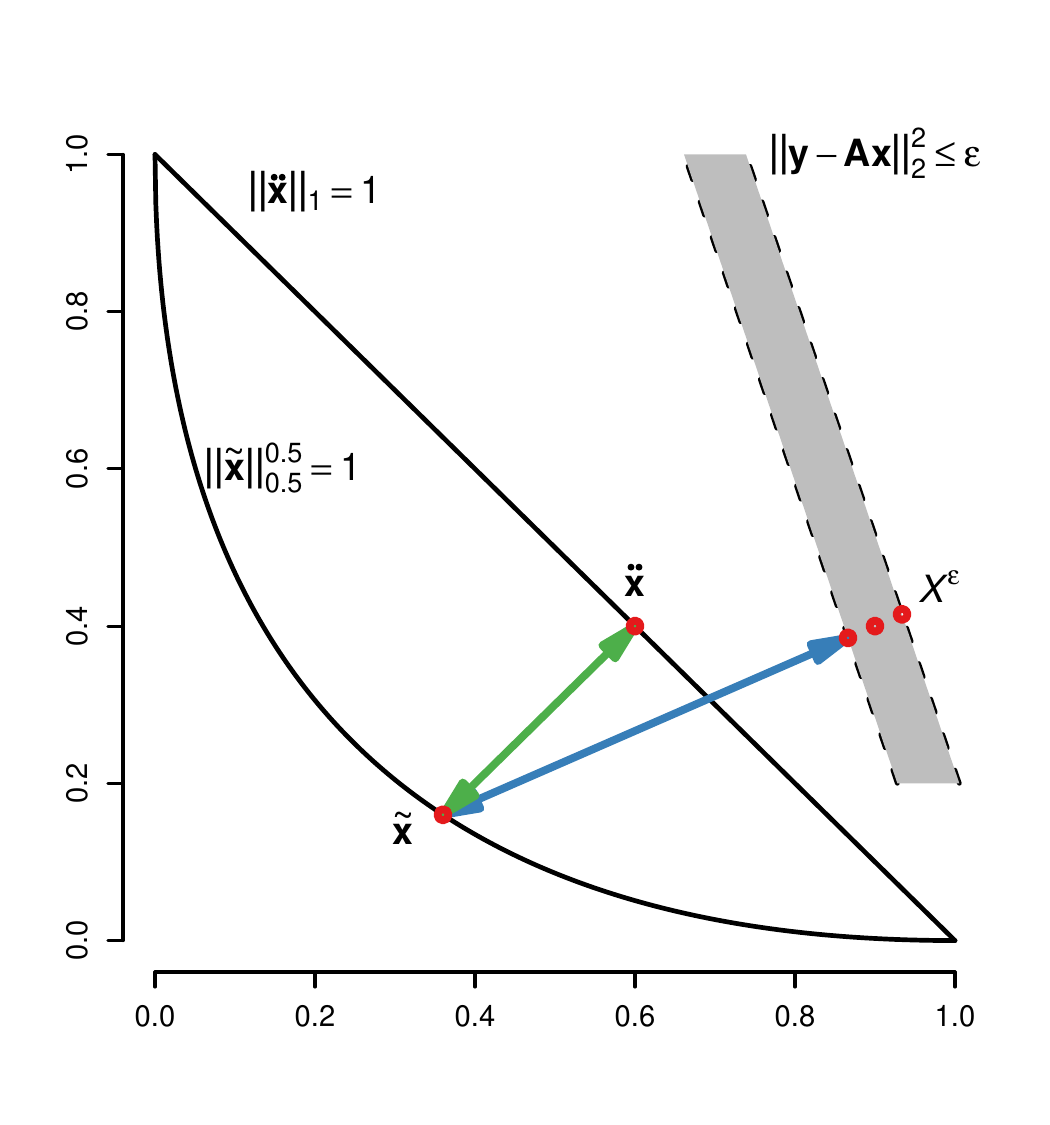}
\caption{The one-to-one mapping in the noisy case: $\mathcal{X}^\epsilon\longleftrightarrow\tilde{\vx}\longleftrightarrow\ddot{\vx}$ when $p=0.5$.}
\label{fig:mapping_noisy}
\end{figure}

Let $\mathcal{X}$ be the set defined in Proposition \ref{proposition:one_to_one_1_noisy}. Specifically, $\mathcal{X}^\epsilon=\mathcal{X}^\epsilon_1\cup\mathcal{X}^\epsilon_2$ and $\mathcal{X}^\epsilon_1\cap\mathcal{X}^\epsilon_2=\emptyset$, where $\mathcal{X}^\epsilon_1$ contains solutions on the boundary of the orthant $\mathbb{O}$, and $\mathcal{X}^\epsilon_2$ contains the rest solutions that are not on the boundary. The solution $\vx$ is then mapped to $\ddot{\vx}$ one by one, producing the corresponding sets $\ddot{\mathcal{X}}_1, \ddot{\mathcal{X}}_2$ on the simplex $\|\ddot{x}\|_1=1$. Similarly we have:
\begin{proposition}
\label{proposition:sparser_solutions_noise}
In the noisy case where $\|\vy-\vA\vx\|_2^2\leq\epsilon$, for every solution $\vx\in\mathcal{X}^\epsilon_2$, there is a solution $\vx^*\in\mathcal{X}^\epsilon_1$ on the boundary of the orthant $\mathbb{O}$ such that $h_p(\vx^*)<h_p(\vx)$ and $h_{p,\alpha}(\vx^*)<h_{p,\alpha}(\vx)$.
\end{proposition}
\begin{proof}
From proposition \ref{proposition:sparser_solutions}, we have that for every $\ddot{\vx}\in\ddot{\mathcal{X}}_2$, there is a $\ddot{\vx}^*\in\ddot{\mathcal{X}}_1$ such that $g(\ddot{\vx}^*)<g(\ddot{\vx})$.

Proposition \ref{proposition:one_to_one_1_noisy} and \ref{proposition:one_to_one_2} guarantee a one to one mapping in $\mathbb{O}$ between $\mathcal{X}^\epsilon$ and $\ddot{\vx}$: $\mathcal{X}^\epsilon\longleftrightarrow\ddot{\vx}$. Since $h_p(\vx)=g(\ddot{\vx})$ and $h_{p,\alpha}(\vx)=g_{\alpha}(\ddot{\vx})$, for every $\vx\in\mathcal{X}^\epsilon_2$, there is an $\vx^*\in\mathcal{X}^\epsilon_1$ such that $h_p(\vx^*)<h_p(\vx)$ and $h_{p,\alpha}(\vx^*)<h_{p,\alpha}(\vx)$.
\end{proof}

We can see that minimizing $h_p(\vx)$ or $h_{p,\alpha}(\vx)$ in the orthant $\mathbb{O}$ also leads to sparser solutions in $\mathcal{X}^\epsilon_1$ in the noisy case.
\end{noisy_r}

\begin{local_minimums}
Here we formally introduce the following Proposition \ref{proposition:local_minimum} on the local minimums of the generalized entropy functions:

\begin{figure}[tbp]
\centering
\includegraphics[height=2.1in]{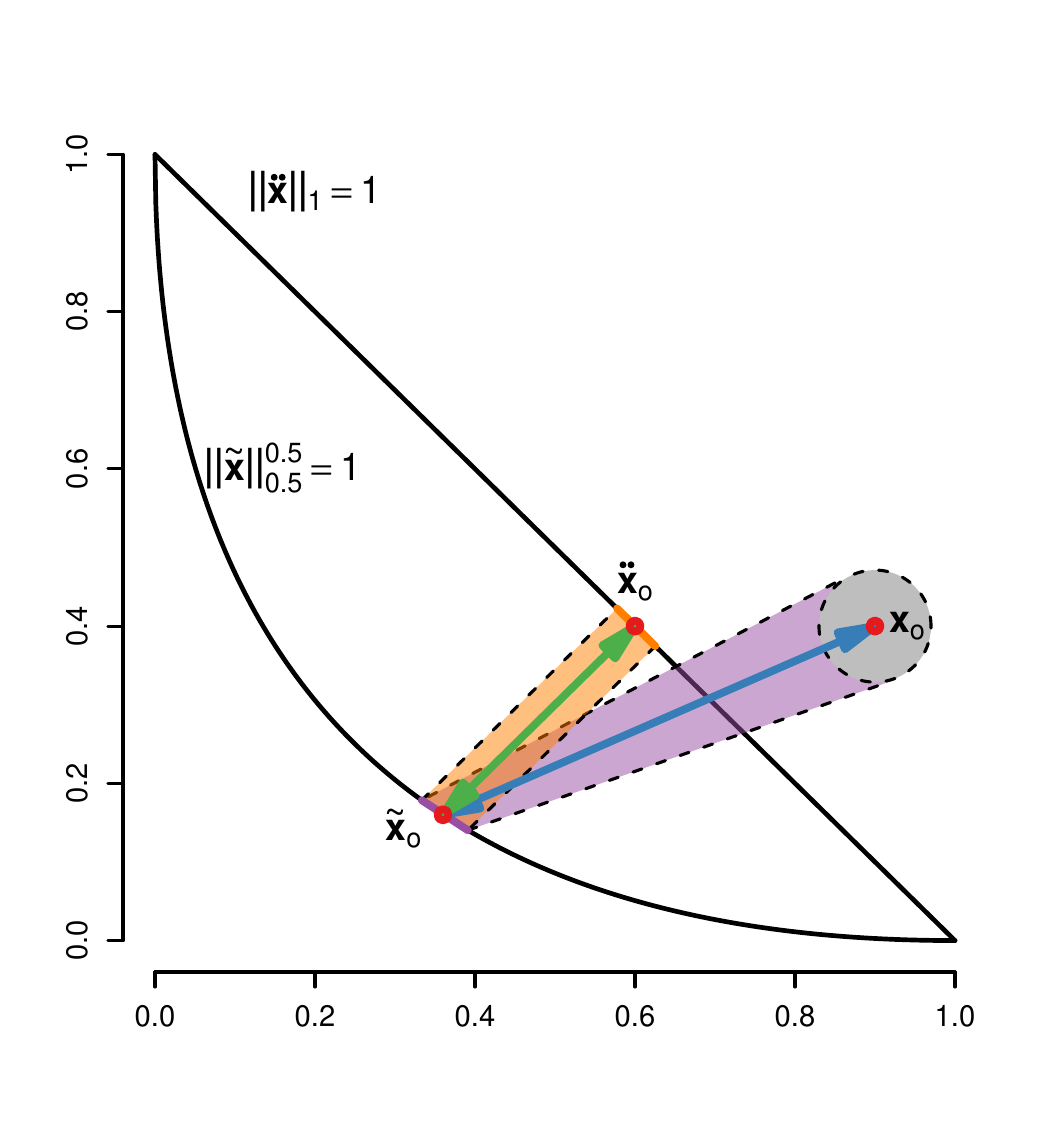}
\caption{The one-to-one mapping: $\rho(\vx_o)\longleftrightarrow\rho(\tilde{\vx}_o)\longleftrightarrow\rho(\ddot{\vx}_o)$ when $p=0.5$.}
\label{fig:mapping_local_minimum}
\end{figure}

\begin{proposition}
\label{proposition:local_minimum}

The local minimums of the generalized entropy functions $h_p(\vx)$ and $h_{p,\alpha}(\vx)$ only occur at the boundaries of each orthant in $\mathbb{R}^N$.
\end{proposition}
\end{local_minimums}
\begin{proof}
Without loss of generality, suppose $\vx_o\in\mathbb{R}^N$ is a local minimum of the generalized entropy function inside some orthant $\mathbb{O}$, as illustrated in Fig. \ref{fig:mapping_local_minimum}. There exists a small neighborhood $\rho(\vx_o)$ surrounding $\vx_o$ such that $\forall \vx\in\rho(\vx_o)$, $h_p(\vx)\geq h_p(\vx_o)$ or $h_{p,\alpha}(\vx)\geq h_{p,\alpha}(\vx_o)$. 

As done in Proposition \ref{proposition:one_to_one_1} and \ref{proposition:one_to_one_1_noisy}, we project $\vx_o$ along with its neighborhood $\rho(\vx_o)$ respectively to $\tilde{\vx}_o$ and $\rho(\tilde{\vx}_o)$ on the sphere $\|\tilde{\vx}\|_p^p=1$ inside the same orthant $\mathbb{O}$:
\begin{align}
\vx_o\longleftrightarrow\tilde{\vx}_o\quad\textnormal{and}\quad\rho(\vx_o)\longleftrightarrow\rho(\tilde{\vx}_o)\,.
\end{align}
$\tilde{\vx}_o$ and $\rho(\tilde{\vx}_o)$ are further projected to $\ddot{\vx}_o$ and $\rho(\ddot{\vx}_o)$ on the simplex $\|\ddot{\vx}\|_1=1$ inside the same orthant $\mathbb{O}$.
\begin{align}
\tilde{\vx}_o\longleftrightarrow\ddot{\vx}_o\quad\textnormal{and}\quad\rho(\tilde{\vx}_o)\longleftrightarrow\rho(\ddot{\vx}_o)\,.
\end{align}
Consequently, $\forall\ddot{\vx}\in\rho(\ddot{\vx}_o)$, we have $h_p(\ddot{\vx})\geq h_p(\ddot{\vx}_o)$ or $h_{p,\alpha}(\ddot{\vx})\geq h_{p,\alpha}(\ddot{\vx}_o)$. $\ddot{\vx}_o$ is also a corresponding local minimum $\ddot{\vx}_o$ on the simplex $\|\ddot{\vx}\|_1=1$ inside the orthant $\mathbb{O}$.

However, we can show that such a local minimum $\ddot{\vx}_o$ does not exist on the simplex $\|\ddot{\vx}\|_1=1$ inside the orthant $\mathbb{O}$:
\begin{enumerate}
\item SEF $h_p(\ddot{\vx})$ with $p>0$: It is strictly concave with respect to $\ddot{\vx}$, there are no local minimums on the simplex $\|\ddot{\vx}\|_1=1$ inside the orthant $\mathbb{O}$, i.e. $\ddot{\vx}_o$ does not exist.
\item REF $h_{p,\alpha}(\ddot{\vx})$ with $p>0$, $\alpha>0$ and $\alpha\neq1$: When $\alpha\in(0,1)$, $h_{p,\alpha}(\ddot{\vx})$ is strictly concave with respect to $\ddot{\vx}$. When $\alpha\in(1,\infty)$, $h_{p,\alpha}(\ddot{\vx})$ is strictly Schur concave with respect to $\ddot{\vx}$ \cite{Bosyk2016}. Hence $\ddot{\vx}_o$ does not exist either.
\end{enumerate}

We can see that there are no local minimums inside each orthant $\mathbb{O}$. Furthermore, the local minimums of $h_p(\ddot{\vx})$ and $h_{p,\alpha}(\ddot{\vx})$ occur at the boundaries of the simplex $\|\ddot{\vx}\|_1=1$. Hence the local minimums of $h_p(\vx)$ and $h_{p,\alpha}(\vx)$ also only occur at the boundary of each orthant $\mathbb{O}$.
\end{proof}

\subsection{Energy-promoting Analysis}
\label{sec:energy_promoting}

The sparsity-promoting property of the various regularizers already encourages the energy of the signal $\|\vx\|_2^2$ to be concentrated in a few significant coefficients. Here we can show that minimizing the generalized entropy functions $h_p(\vx)$ and $h_{p,\alpha}(\vx)$ further \emph{promotes} multiple high-energy coefficients while \emph{suppressing} the rest low-energy coefficients within the recovered solution. 

\begin{sef_ene}
The energy-promoting property can be best illustrated from an optimization perspective. Take the SEF $h_p(\vx)$ for example, should we use the subderivative descent to minimize it, the solution $\vx$ is updated as follows:
\begin{align}
\label{eq:gd_update_shannon}
\begin{split}
\vx &= \vx-\eta\cdot\partial h_p(\vx)\\
&=\vx-\eta\cdot\textnormal{sign}(\vx)\circ\nabla h_p(|\vx|)\,,
\end{split}
\end{align}
where $\eta>0$ is some suitable step size, and $\nabla h_p(|\vx|)$ is the gradient with respect to the magnitude $|\vx|$:
\begin{align}
\label{eq:derivative_shannon_en_fun}
\begin{split}
&\frac{\partial h_p(\vx)}{\partial |x_i|}=\frac{\partial}{\partial |x_i|}\left(-\frac{\sum_{i=1}^N|x_i|^p\log|x_i|^p}{\|\vx\|_p^p}+\log\|\vx\|_p^p\right)\\
&=-\frac{1}{\|\vx\|_p^p}\left(p|x_i|^{p-1}\log |x_i|^p+p|x_i|^{p-1}\right)\\
&\quad+\frac{p|x_i|^{(p-1)}\sum_{l=1}^N |x_l|^p\log|x_l|^p}{\|\vx\|_p^{2p}}+\frac{p|x_i|^{p-1}}{\|\vx\|_p^p}\\
&=-\frac{p|x_i|^{(p-1)}\log |x_i|^p}{\|\vx\|^p_p}+\frac{p|x_i|^{(p-1)}\sum_{l=1}^N |x_l|^p\log|x_l|^p}{\|\vx\|_p^{2p}}.
\end{split}
\end{align}
It's easy to verify the following remark:
\begin{remark}
Let $\nu = \exp\left(\frac{\sum_l|x_l|^p\log|x_l|^p}{p\|\vx\|_p^p}\right),\nu>0$, we have:
\begin{align}
\label{eq:pd_nu_sef}
\frac{\partial h_p(\vx)}{\partial |x_i|}\left\{
\begin{array}{l}
<0 \\
=0 \\
>0
\end{array} \quad 
\begin{array}{l}
\textnormal{if $|x_i|>\nu$}\\
\textnormal{if $|x_i|=\nu$}\\
\textnormal{if $|x_i|<\nu$}\,.
\end{array}
\right.
\end{align}
\end{remark}

\begin{figure}[tbp]
\centering
\includegraphics[height=1.5in]{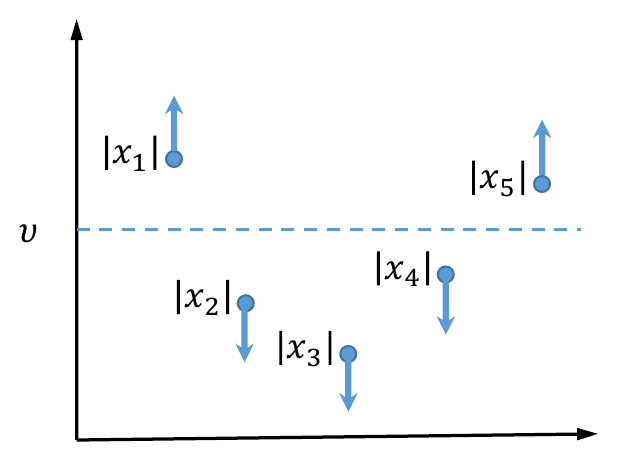}
\caption{When the magnitude of the $i$-th entry $|x_i|>\nu$, the update in (\ref{eq:gd_update_shannon}) makes $|x_i|$ larger. When $|x_i|<\nu$, the update in (\ref{eq:gd_update_shannon}) makes $|x_i|$ smaller.}
\label{fig:energy_concen}
\end{figure}

As illustrated in Fig. \ref{fig:energy_concen}, for the entries with relatively large magnitudes $|x_i|>\nu$, we can see that the update in (\ref{eq:gd_update_shannon}) makes their magnitudes even larger. Conversely, if the entries have relatively small magnitudes $|x_i|<\nu$, the update makes their magnitudes even smaller. In this way minimizing $h_p(\vx)$ promotes large-magnitude entries while suppressing small-magnitude entries in the recovered signal.
\end{sef_ene}

\begin{ref_ene}
For the REF $h_{p,\alpha}(\vx)$, the derivative with respect to the magnitude $|\vx|$ is:
\begin{align}
\label{eq:derivative_renyi_en_fun}
\begin{split}
\frac{\partial h_{p,\alpha}(\vx)}{\partial |x_i|}&=\frac{1}{1-\alpha} \times \frac{1}{\sum_{l=1}^N\left(\frac{|x_l|}{\|\vx\|_p}\right)^{p\alpha}} \times \frac{p\alpha}{\|\vx\|_p^{p(\alpha+1)}}\\
&\quad\quad\times\left[|x_i|^{p\alpha-1}\|\vx\|_p^p-|x_i|^{p-1}\|\vx\|_{p\alpha}^{p\alpha}\right]\,.
\end{split}
\end{align}
Similarly we have the following remark:
\begin{remark}
Let $\nu=\exp\left(\frac{1}{p\alpha-p}\log\frac{\|\vx\|_{p\alpha}^{p\alpha}}{\|\vx\|_p^p}\right), \nu>0$, we have:
\begin{align}
\label{eq:pd_nu_ref}
\frac{\partial h_{p,\alpha}(\vx)}{\partial |x_i|}\left\{
\begin{array}{l}
<0 \\
=0 \\
>0
\end{array} \quad 
\begin{array}{l}
\textnormal{if $|x_i|>\nu$}\\
\textnormal{if $|x_i|=\nu$}\\
\textnormal{if $|x_i|<\nu$}\,.
\end{array}
\right.
\end{align}
\end{remark}
We can see that minimizing $h_{p,\alpha}(\vx)$ also promotes large-magnitude entries while suppressing small-magnitude entries in the recovered signal.
\end{ref_ene}

As discussed in Appendix \ref{app:energy_promoting_other}, we can see that minimizing the other regularizers introduced in section \ref{sec:sparsity_regularizers} do not have such energy-promoting behaviors.

\vspace{1mm}
\subsubsection{Advantage in SRC}
\label{sec:adv_src}
This energy-promoting behavior becomes handy in applications such as sparse representation classification (SRC) \cite{SRC_face09,SRC_CVPR10}. Specifically, suppose the subspace of interest is spanned by the columns of $\vA=[\vA_1\,\vA_2 \cdots \vA_C\,\vI]$, where $\vA_i$ is a block matrix whose columns are the collected samples belonging to the $i$-th class, and $\vI$ is the identity matrix used as the dictionary for the noise. The $\vx$ can be written accordingly as $\vx^\textnormal{T}=[\vx_1^\textnormal{T}\,\,\vx_2^\textnormal{T}\,\,\cdots\vx_C^\textnormal{T}\,\,\vx_{\vI}^\textnormal{T}]$, where $\vx_i$ contains the coefficients corresponding to the $i$-th class, and $\vx_{\vI}$ is the noise. Additionally, we assume the block matrix $\vA_i$ satisfies the following restricted isometry property for some $\delta(i)\in(0,1)$ \cite{RIP08}:
\begin{align}
\sqrt{1-\delta(i)}\|\vx_i\|_2\leq\|\vA_i\vx_i\|_2\leq\sqrt{1+\delta(i)}\|\vx_i\|_2\,.
\end{align}

After the solution $\vx$ that satisfies the constraint $\|\vy-\vA\vx\|_2\leq\sqrt{\epsilon}$ is computed, the following residues are computed for every class:
\begin{align}
\sigma_i = \|\vy-\vx_{\vI}-\vA_i\vx_i\|_2,\quad\forall i\in[1,\cdots,C]\,.
\end{align}
The class label of $\vy$ is then:
\begin{align}
c=\arg\textstyle\min_i\,\,\sigma_i\,.
\end{align}
Without loss of generality, suppose $\vy$ belongs to the $1$st class, and we are comparing $\sigma_1$ and $\sigma_2$. Let $\overline{\vA_{1,2}}$ denote the matrix obtained after removing $\vA_1$ and $\vA_2$ from $\vA$, while $\overline{\vx_{1,2}}$ denotes the vector obtained after removing $\vx_1$ and $\vx_2$ from $\vx$. We then have:
\begin{align}
\|\vy-\vx_{\vI}-\vA_1\vx_1-\vA_2\vx_2-\overline{\vA_{1,2}}\overline{\vx_{1,2}}\|_2\leq\sqrt{\epsilon}\,.
\end{align}
Using the reverse triangle inequality, we have:
\begin{align}
&\|\vy-\vx_{\vI}-\vA_1\vx_1\|_2-\|\vA_2\vx_2+\overline{\vA_{1,2}}\overline{\vx_{1,2}}\|_2\leq\sqrt{\epsilon}\\[1ex]
\label{eq:sigma_1_ub}
\begin{split}
&\sigma_1=\|\vy-\vx_{\vI}-\vA_1\vx_1\|_2\\
&\quad\leq\|\vA_2\vx_2\|_2+\|\overline{\vA_{1,2}}\overline{\vx_{1,2}}\|_2+\sqrt{\epsilon}\\
&\quad\leq\sqrt{1+\delta(2)}\|\vx_2\|_2+\textstyle\sum_{i\neq\{1,2\}}\sqrt{1+\delta(i)}\|\vx_i\|_2+\sqrt{\epsilon}\,.
\end{split}
\end{align}
Using the reverse triangle inequality, we have:
\begin{align}
&\|\vy-\vx_{\vI}-\vA_2\vx_2\|_2-\|\vA_1\vx_1+\overline{\vA_{1,2}}\overline{\vx_{1,2}}\|_2\geq-\sqrt{\epsilon}\\[1ex]
\label{eq:sigma_2_lb}
\begin{split}
&\sigma_2=\|\vy-\vx_{\vI}-\vA_2\vx_2\|_2\\
&\quad\geq\|\vA_1\vx_1+\overline{\vA_{1,2}}\overline{\vx_{1,2}}\|_2-\sqrt{\epsilon}\\
&\quad\geq\|\vA_1\vx_1\|_2-\|\overline{\vA_{1,2}}\overline{\vx_{1,2}}\|_2-\sqrt{\epsilon}\\
&\quad\geq\sqrt{1-\delta(1)}\|\vx_1\|_2-\textstyle\sum_{i\neq\{1,2\}}\sqrt{1+\delta(i)}\|\vx_i\|_2-\sqrt{\epsilon}\,.
\end{split}
\end{align}
We can see that \eqref{eq:sigma_1_ub} gives an upper bound on $\sigma_1$, and we would like to keep it small. Hence $\|\vx_i\|_2$, $\forall i\neq 1$ should be small. While \eqref{eq:sigma_2_lb} gives a lower bound on $\sigma_2$, and we would like to keep it large. Hence $\|\vx_1\|_2$ should be large, while $\|\vx_i\|$, $\forall i\neq\{1,2\}$ should be small. In other words, we would like to promote the energy of $\vx_1$ and suppress the energy of $\vx_i$, $\forall i\neq 1$ at the same time. The energy-promoting property of the generalized entropy functions is well suited for this purpose.

\section{Entropy Function Minimization}
Here we present the algorithm to perform the sparse signal recovery tasks formalized in \eqref{eq:sefm_ssr},\eqref{eq:refm_ssr}. The data fidelity term $f(\vx)$ is convex and smooth, the sparsity regularizers $h_p(\vx), h_{p,\alpha}(\vx)$ are nonconvex and nonsmooth. FISTA \cite{FISTA09} is adapted as the accelerated inexact proximal gradient method \cite{APG2015} to minimize the nonconvex objective functions.

\subsection{Inexact Proximal Gradient Method}
Using the generalized Shannon entropy function $h_p(\vx)$ as an example, we first introduce the inexact proximal gradient method to solve the nonconvex problem \eqref{eq:sefm_ssr}:
\vspace{0.1cm}
\begin{first_step}
Proximal regularization using the data fidelity term $f(\vx)=\|\vy-\vA\vx\|_2^2$.

In the $(t+1)$-th iteration, we use the proximal regularization \cite{Martinet1970} of $f(\vx)$ at $\vx^{(t)}$, as done in \cite{FISTA09}:
\begin{align}
\label{eq:proximal_data_fidelity}
\begin{split}
f(\vx)&= \|\vy-\vA\vx\|_2^2\\
&\leq f(\vx^{(t)})+\left<\vx-\vx^{(t)}, \nabla f(\vx^{(t)})\right>+\frac{\kappa}{2}\left\|\vx-\vx^{(t)}\right\|_2^2\\
&= o(\vx^{(t)}) + \frac{\kappa}{2}\left\|\vx-\left(\vx^{(t)}-\frac{1}{\kappa}\nabla f(\vx^{(t)})\right)\right\|_2^2\,,
\end{split}
\end{align}
where $o(\vx^{(t)})$ is a relative constant depending on the previous $t$-th iteration's solution $\vx^{(t)}$ and can thus be ignored; $\nabla f(\vx^{(t)}) = 2(\vA^\textrm{T}\vA\vx^{(t)}-\vA^\textrm{T}\vy)$; the $\kappa$ is the Lipschitz constant of the gradient $\nabla f(\vx^{(t)})$ \cite{Eriksson_appliedmathematics}. The smallest value $\kappa$ can take is \emph{twice} the largest eigenvalue of $\vA^\textrm{T}\vA$, this is to ensure that $f(\vx)$ in \eqref{eq:proximal_data_fidelity} is upper-bounded by the proximal regularization at the right-hand side. When $\kappa$ is unknown or difficult to compute, we can use the backtracking strategy describe in \cite{FISTA09} to find it.

Plug \eqref{eq:proximal_data_fidelity} into \eqref{eq:sefm_ssr}, we need to solve the following proximal operator of the nonconvex function $h_p(\vx)$:
\begin{align}
\label{eq:proximal_shannon_en_fun}
\begin{split}
&\textnormal{prox}_{\frac{\lambda}{\kappa}h_p}\left(\vx^{(t)}-\frac{1}{\kappa}\nabla f(\vx^{(t)})\right) = \arg\min_{\vx} R_p(\vx)\\
&=\arg\min_{\vx}\frac{\kappa}{2}\left\|\vx-\left(\vx^{(t)}-\frac{1}{\kappa}\nabla f(\vx^{(t)})\right)\right\|_2^2 + \lambda\,\, h_p(\vx)\,.
\end{split}
\end{align}
\end{first_step}

\vspace{0.1cm}
\begin{second_step}
The nonconvex proximal operator \eqref{eq:proximal_shannon_en_fun} is inexactly solved.

Since \eqref{eq:proximal_shannon_en_fun} is nonconvex, it is still an open-problem to obtain closed-form or exact solution \cite{APG2015, NoncxProxSum2016, Noncxprox2017}. Our goal here is to approximately solve it so that the solution $\vr$:
\begin{align}
\label{eq:min_constraint_r}
R_p(\vr)&\leq R_p(\vx^{(t)})\,.
\end{align}
As shown in Appendix \ref{app:mono_min}, this would ensure the solution converges to a critical point.

The nonconvex function $h_p(\vx)$ is replaced with its first-order approximation:
\begin{align}
\label{eq:approx_shannon_entropy}
\begin{split}
h_p(\vx)&\approx h_p(|\vx^{(t)}|) + \left<|\vx|-|\vx^{(t)}|,\nabla h_p(|\vx^{(t)}|)\right>\,,
\end{split}
\end{align}
where $\nabla h_p(|\vx|)$ is the gradient with respect to the modulus $|x_i|$ given in \eqref{eq:derivative_shannon_en_fun}. When computing $\nabla h_p(|x_i^{(t)}|)$, we add a small positive value $\epsilon=10^{-12}$ to $|x_i^{(t)}|$ to avoid $\log 0$. Ignoring the relative constant terms in \eqref{eq:approx_shannon_entropy} that depend only on $\vx^{(t)}$, the proximal operator \eqref{eq:proximal_shannon_en_fun} can be approximated:
\begin{align}
\label{eq:shannon_en_approx_secondstep}
\begin{split}
\vr=\arg\min_{\vx}\,&\frac{\kappa}{2}\left\|\vx-\vs^{(t)}\right\|_2^2 + \lambda \left<|\vx|,\nabla h_p(|\vx^{(t)}|)\right>\,,
\end{split}
\end{align}
where $\vs^{(t)}=\vx^{(t)}-\frac{1}{\kappa}\nabla f(\vx^{(t)})$. This is a simple reweighted $l_1$-norm minimization problem that consists of a series of independent one-dimensional problems. The solution $r_i$ to \eqref{eq:shannon_en_approx_secondstep} can be obtained using the soft shrinkage operator:
\begin{align}
\label{eq:rw_shannon_en_fun}
r_i&=\mathlarger\Gamma_{\frac{\lambda}{\kappa}\nabla h_p(|x_i^{(t)}|)}\left(s_i^{(t)}\right)\,,
\end{align}
where $\Gamma_\tau(\cdot)$ is defined as follows:
\begin{align}
\label{eq:shrinkage}
\Gamma_\tau(x)=\left\{
\begin{array}{l}
0 \\
(|x|-\tau)\cdot \textrm{sign}(x)
\end{array} \quad 
\begin{array}{l}
\textrm{if $|x|\leq\tau$}\\
\textrm{if $|x|>\tau$}\,.
\end{array}
\right.
\end{align}
Conventionally, the soft shrinkage operator solves a convex problem and requires the threshold $\tau$ to be positive. However, the gradient $\nabla h_p(|\vx^{(t)}|)$ could be negative, which makes \eqref{eq:shannon_en_approx_secondstep} nonconvex. In Appendix \ref{app:gsst} we show that the global optimal solution can still be obtained using the soft thresholding operator in (\ref{eq:shrinkage}), yet with a different derivation process. 

We further update $\vr$ as follows to ensure \eqref{eq:min_constraint_r} holds:
\begin{align}
\vr=\vx^{(t)}\quad\textrm{if $R_p(\vr)> R_p(\vx^{(t)})$}\,.
\end{align}
\end{second_step}

\begin{figure*}[tbp]
\centering
\subfigure{
\label{fig:tune_p}
\includegraphics[height=1.7in]{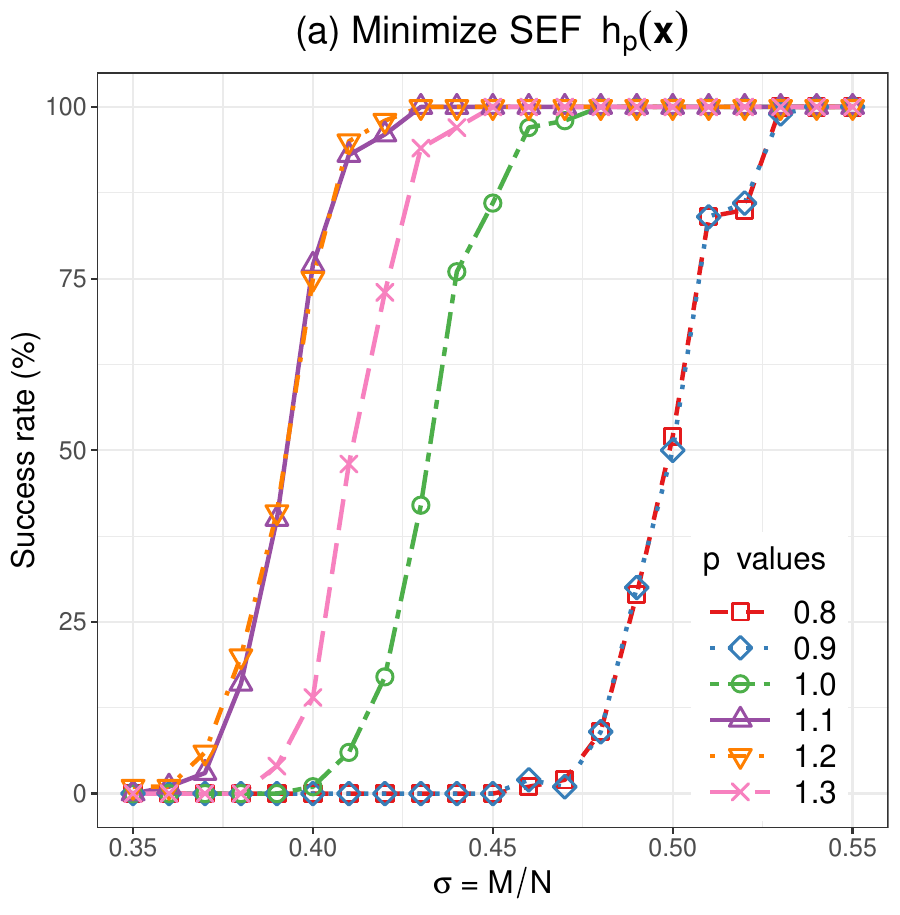}}
\subfigure{
\label{fig:tune_p_alpha_09_11}
\includegraphics[height=1.7in]{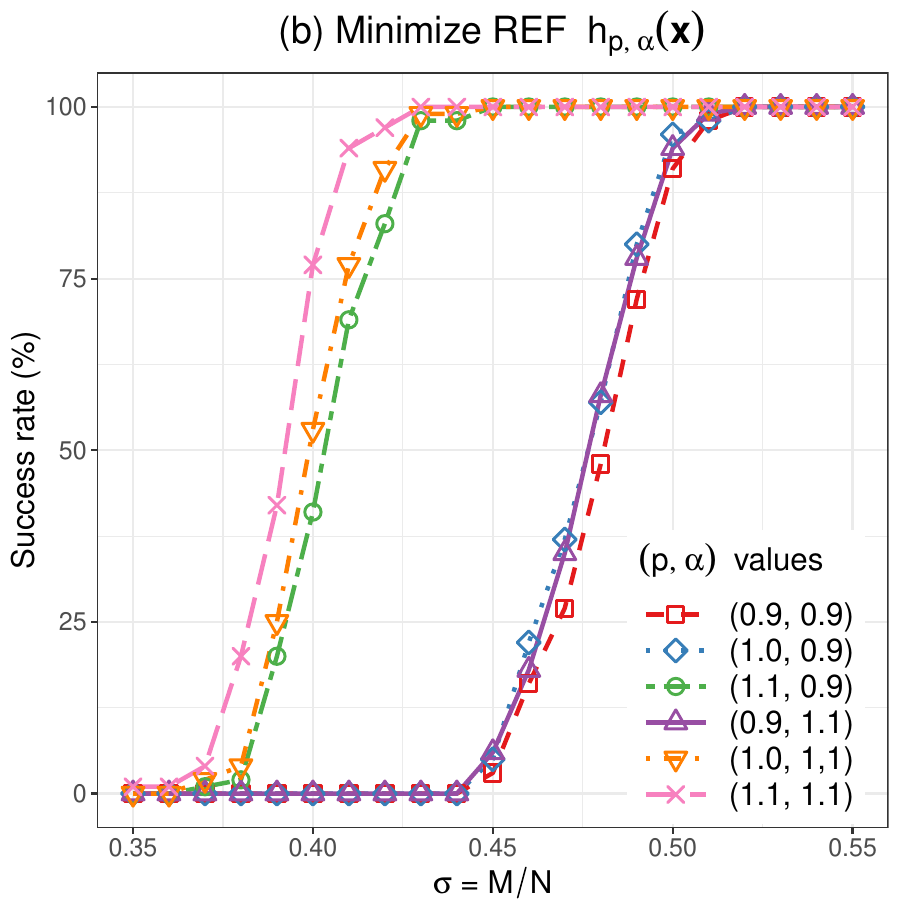}}
\subfigure{
\label{fig:init_tune_p}
\includegraphics[height=1.7in]{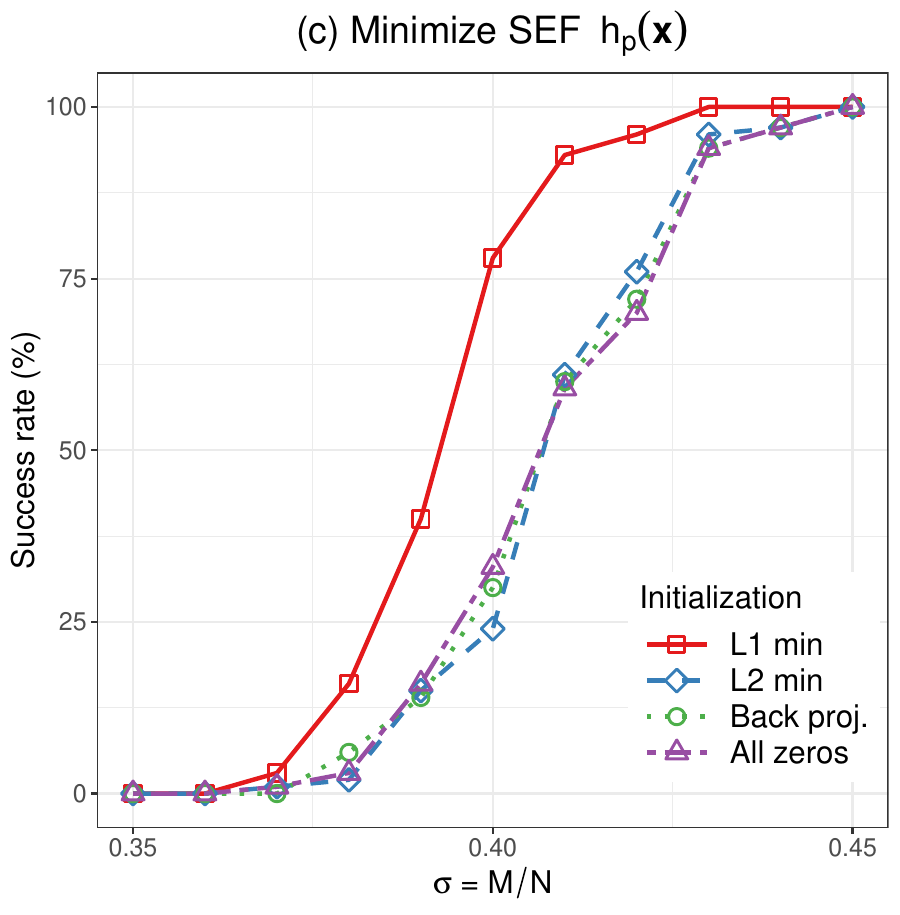}}
\subfigure{
\label{fig:init_tune_p_alpha}
\includegraphics[height=1.7in]{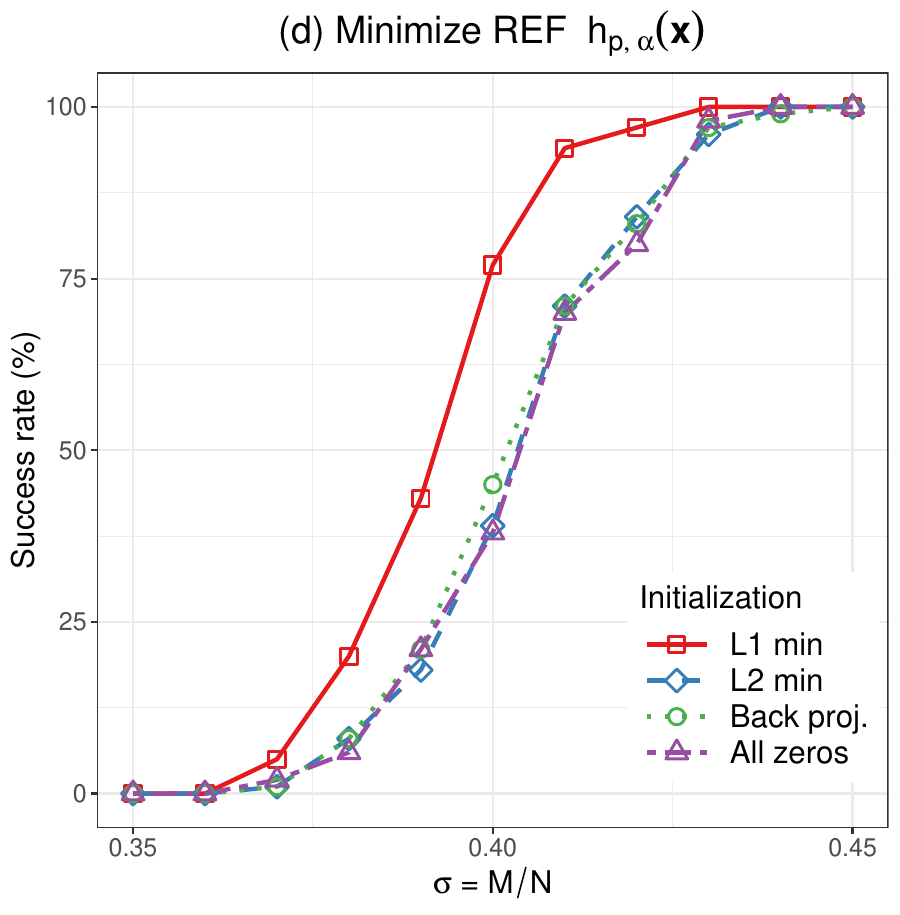}}
\caption{\ref{fig:tune_p} and \ref{fig:tune_p_alpha_09_11}: the success rates of the recovery experiments with different parameter choices; \ref{fig:init_tune_p} and \ref{fig:init_tune_p_alpha} the success rates of the recovery experiments with different initialization strategies.}
\label{fig:entropy_fun_p_val_init}
\end{figure*}

\subsection{Accelerated Inexact Proximal Gradient Method}
\label{sec:acc_ipgm}
The proximal gradient method \cite{Martinet1970, Proximal13} was proposed to solve convex optimization problems, and the error from the optimum objective function is guaranteed to be within $O(\frac{1}{t})$ after $t$ iterations. Beck and Teboulle \cite{FISTA09,Beck2009FastGA} later proposed the accelerated proximal gradient method, a.k.a. FISTA, that has a convergence rate of $O(\frac{1}{t^2})$. Li and Lin \cite{APG2015} later extended the accelerated proximal gradient (APG) to handle nonconvex and nonsmooth problems. In this case, the APG is guaranteed to converge to a critical point of the objective function \cite{APG2015}. Empirically, the APG could speed up the convergence by a great deal. However, it is still an open problem and a nontrival task to perform convergence rate analysis for general nonconvex objective functions \cite{APG2015, NoncxProxSum2016, Noncxprox2017}. 

Here we adapt the nonconvex acceleration approach by \cite{APG2015} to solve \eqref{eq:sefm_ssr}. Let $F_p(\vx)=f(\vx)+\lambda\,\,h_p(\vx)$, we have:
\begin{align}
\label{eq:update_acceleration_s}
&\vu^{(t)} = \vx^{(t)}+\frac{k^{(t-1)}}{k^{(t)}}(\vz^{(t)}-\vx^{(t)})+\frac{k^{(t-1)}-1}{k^{(t)}}(\vx^{(t)}-\vx^{(t-1)})\\
&\vz^{(t+1)} = \textnormal{prox}_{\frac{\lambda}{\kappa}h_p}\left(\vu^{(t)}-\frac{1}{\kappa}\nabla f(\vu^{(t)})\right)\\
&\vv^{(t+1)} = \textnormal{prox}_{\frac{\lambda}{\kappa}h_p}\left(\vx^{(t)}-\frac{1}{\kappa}\nabla f(\vx^{(t)})\right)\\
&k^{(t+1)}=\frac{1+\sqrt{4\left(k^{(t)}\right)^2+1}}{2}\\
\label{eq:update_acceleration_x}
&\vx^{(t+1)}=\left\{
\begin{array}{l}
\vz^{(t+1)}\\
\vv^{(t+1)}
\end{array}\quad
\begin{array}{l}
\textnormal{if $F_p(\vz^{(t+1)})\leq F_p(\vv^{(t+1)})$}\\
\textnormal{otherwise}\,.
\end{array}
\right.
\end{align}

The iterations are stopped either when the maximum number of iterations is reached or when $\vx^{(t+1)}$ reaches convergence, i.e. $\frac{\|\vx^{(t+1)}-\vx^{(t)}\|_2}{\|\vx^{(t+1)}\|_2}\leq\tau_c$. The convergence threshold $\tau_c$ is usually chosen to be $0<\tau_c\leq 10^{-3}$.

Finally, the proposed entropy function minimization approach is summarized in Algorithm \ref{alg:entropy_function_fista}:

\begin{algorithm}[tbp]
\caption{Sparse signal recovery via generalized entropy function minimization}
\label{alg:entropy_function_fista}
\begin{algorithmic}[1]
\REQUIRE $\{\vy, \vA\}, \lambda, \kappa, p, \tau_c, T, k^{(1)}=1, k^{(0)}=0$
\STATE Initialize $\vx^{(0)}$ with the solution from $l_1$-minimization;
\STATE Set $\vz^{(1)}=\vx^{(1)}=\vx^{(0)}$;
\FOR{$t=\{1,\cdots,T\}$}
    \STATE Update $\vu$, $\vz$, $\vv$, $k$, $\vx$ as in \eqref{eq:update_acceleration_s}-\eqref{eq:update_acceleration_x};
	\IF {$\frac{\left\|\vx^{(t+1)}-\vx^{(t)}\right\|_2}{\|\vx^{(t+1)}\|_2}\leq\tau_c$}
		\STATE Set $\hat{\vx}=\vx^{(t+1)}$ and {\bfseries break}.
	\ENDIF
\ENDFOR
\STATE {\bfseries Return} Output $\hat{\vx}$;
\end{algorithmic}
\end{algorithm}

\subsection{Regularization Parameter}
\label{sec:regularization_para}
Naturally, choosing a proper regularization parameter $\lambda$ is the key to the success of sparse signal recovery. 
\subsubsection{Noiseless recovery}
The solution can be obtained by solving a series of minimization problems characterized by a decreasing $\lambda$ sequence: $\lambda\rightarrow0$ \cite{L1review13}. Take the Shannon entropy function $h_p(\vx)$ for example, we have the following:
\begin{algorithmic}[1]
\STATE Start with a relatively large $\lambda_0$, and set $\vx^{\lambda_0}$ to the $l_1$-minimization solution.
\FOR{$l=\{1,\cdots,T\}$}
	\STATE Use $\vx^{\lambda_{l-1}}$ as the initializer, solve the minimization problem characterized by $\lambda_l$: 
	\begin{align}
	\vx^{\lambda_l}=\arg\min_{\vx}\,\|\vy-\vA\vx\|_2^2+\lambda_l\,\,h_p(\vx)\,.
	\end{align}
	\STATE Update $\lambda_{l+1}=\rho\cdot\lambda_l$.
	\STATE \textbf{if} $\frac{\left\|\vx^{\lambda_{l+1}}-\vx^{\lambda_l}\right\|_2}{\|\vx^{\lambda_{l+1}}\|_2}\leq\tau_c$ \textbf{then} break.
\ENDFOR
\end{algorithmic}
To ensure the best performance, $\rho$ is chosen in $[0.9,1)$. 
\subsubsection{Noisy recovery}
The optimal $\lambda$ depends on the noise level and is usually unknown. An optimal fixed $\lambda$ can be pre-tuned on some development set and later used in practice.

\subsection{Constructing Suitable Entropy Functions}
\label{sec:tune_entropy_fun}
The choice of different $p$ and $\alpha$ values also greatly affects the recovery performance. 
\begin{itemize}
\item For the SEF $h_p(\vx)$, where $p>0$. The optimal $p$ value is usually around $1$. In practice, we can focus on tuning $p$ in the range $[0.8,\,1.3]$.
\item For the REF $h_{p,\alpha}(\vx)$, where $p>0$ and $0<\alpha\neq 1$. Both optimal $p$ and $\alpha$ values are usually around $1$. Empirically, we can focus on tuning $p$ in the range $[0.8,\,1.3]$, and $\alpha$ in the range $[0.9,\,1)\cup(1,\,1.1]$.
\end{itemize}

As an example, here we compare the noiseless recovery performances of different entropy functions. For the experiments, $100$ random sparse signals $\vx$ of length $N=1000$ and sparsity level $S=200$ are generated: the nonzero entries are independently and identically drawn from $\mathcal{N}(0,1)$. The sensing matrices $\vA$s are random Gaussian matrices with i.i.d entries following $\mathcal{N}(0,1)$, with the columns normalized and centralized. The number of measurements $M$ is varied in $[350,\,550]$, and Algorithm \ref{alg:entropy_function_fista} is used to obtain the recovered signal $\hat{\vx}$. The recovery is considered to be a success if $\frac{\|\hat{\vx}-\vx\|_2}{\|\vx\|_2}<10^{-3}$. The success rates out of the $100$ recoveries for different choices of $M,p,\alpha$ are shown in Fig. \ref{fig:entropy_fun_p_val_init}.

\subsection{Initialization Strategy}
\label{sec:initialization}
A proper initialization is needed to ensure good performances for the nonconvex problems. Following the same experimental setup as section \ref{sec:tune_entropy_fun}, we choose $p=1.1$ for the SEF $h_p(\vx)$, and $(p=1.1,\, \alpha=1.1)$ for the REF $h_{p,\alpha}(\vx)$. We compare the following initialization strategies:
\begin{enumerate}
\item The $l_1$-minimization solution.
\item The $l_2$-minimization solution, i.e. the Moore-Penrose pseudo-inverse of $\vy$: $\vA^\textnormal{T}(\vA\vA^\textnormal{T})^{-1}\vy$.
\item Back projection: $\vA^\textnormal{T}\vy$.
\item All zeros: $[0,\cdots,0]^\textnormal{T}$.
\end{enumerate}
The success rates using the above initializers are shown in Fig. \ref{fig:entropy_fun_p_val_init}. We can see that the solution from $l_1$-norm minimization usually leads to the best performances. It is thus used as the initializer in the proposed Algorithm \ref{alg:entropy_function_fista}. 

\begin{figure*}[htbp]
\centering
\subfigure{
\label{fig:entropy_ptc_show}
\includegraphics[height=2.15in]{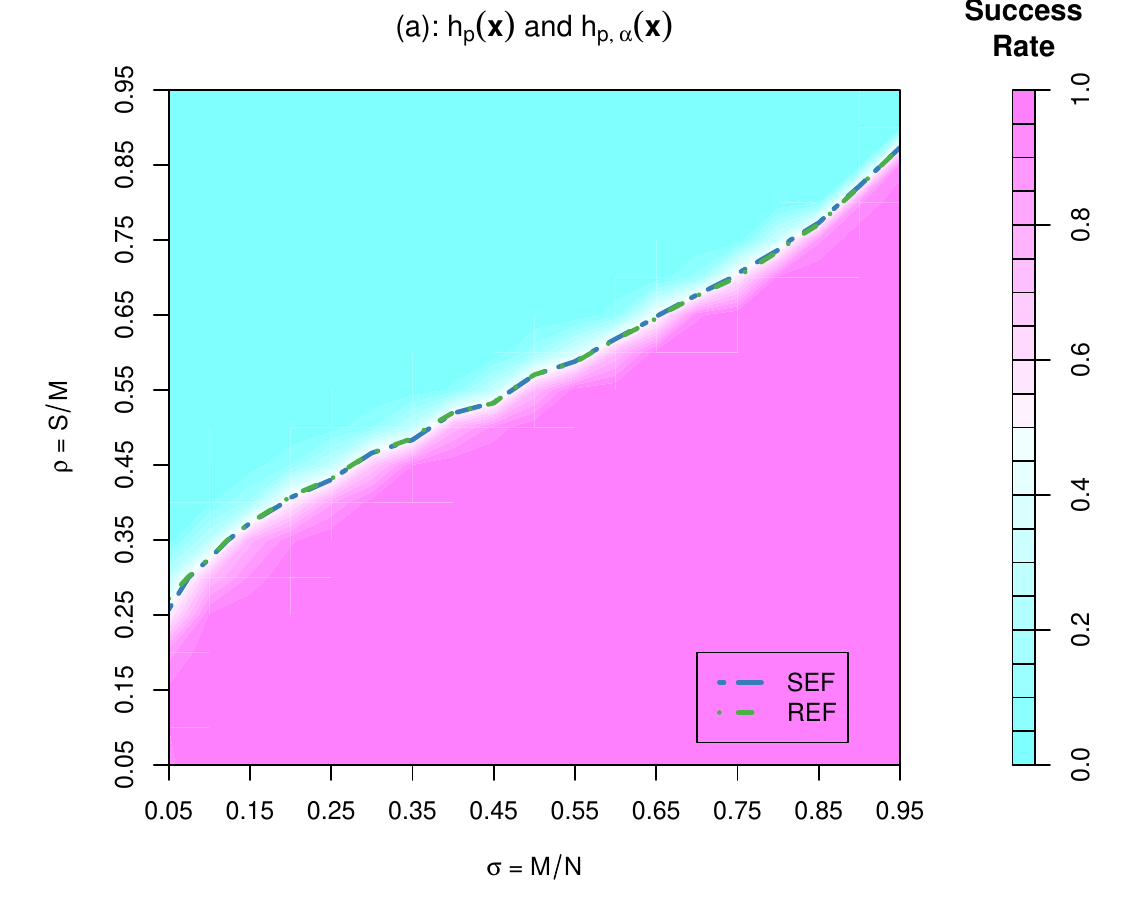}}
\subfigure{
\label{fig:ptc_noiseless_sparse}
\includegraphics[height=2.15in]{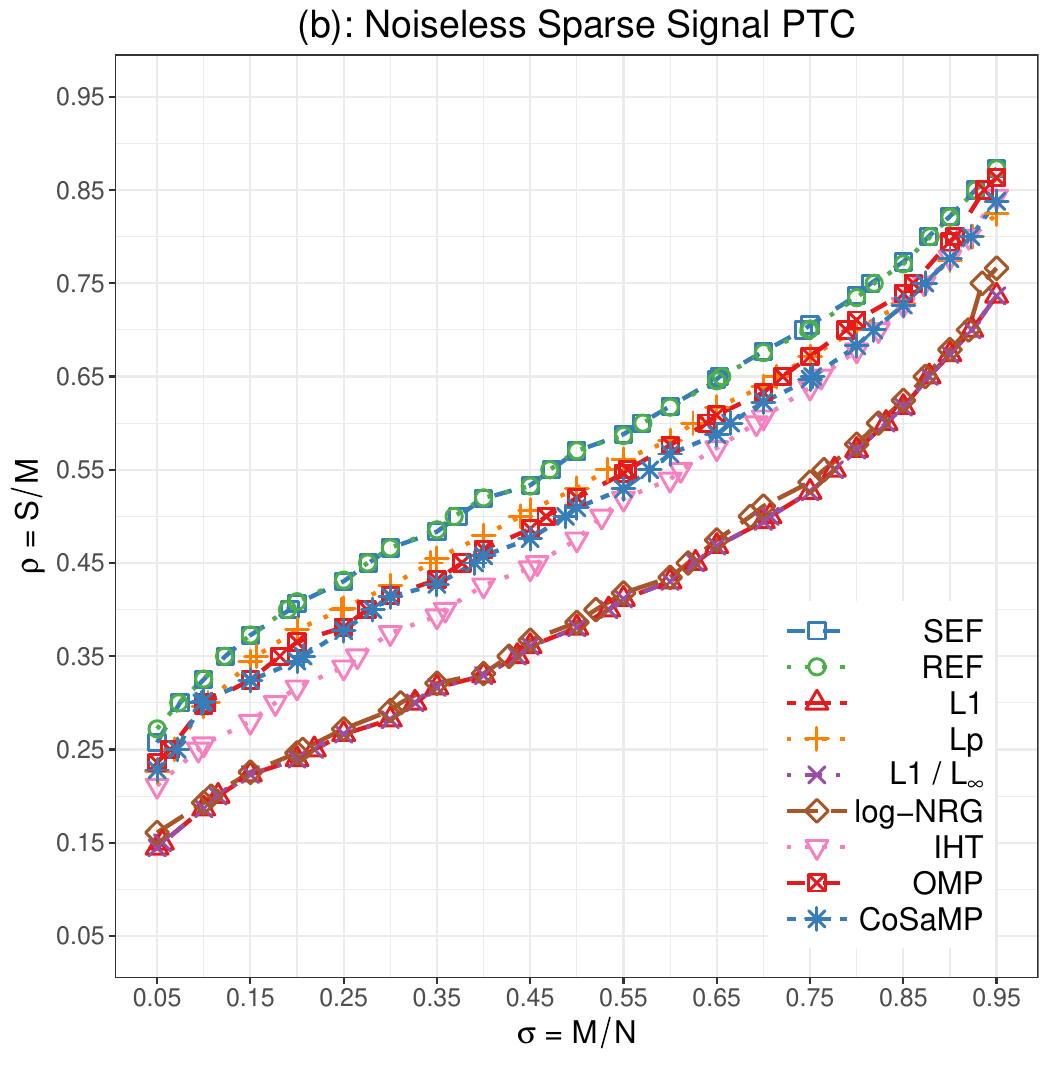}}
\subfigure{
\label{fig:snr_noisy_sparse}
\includegraphics[height=2.15in]{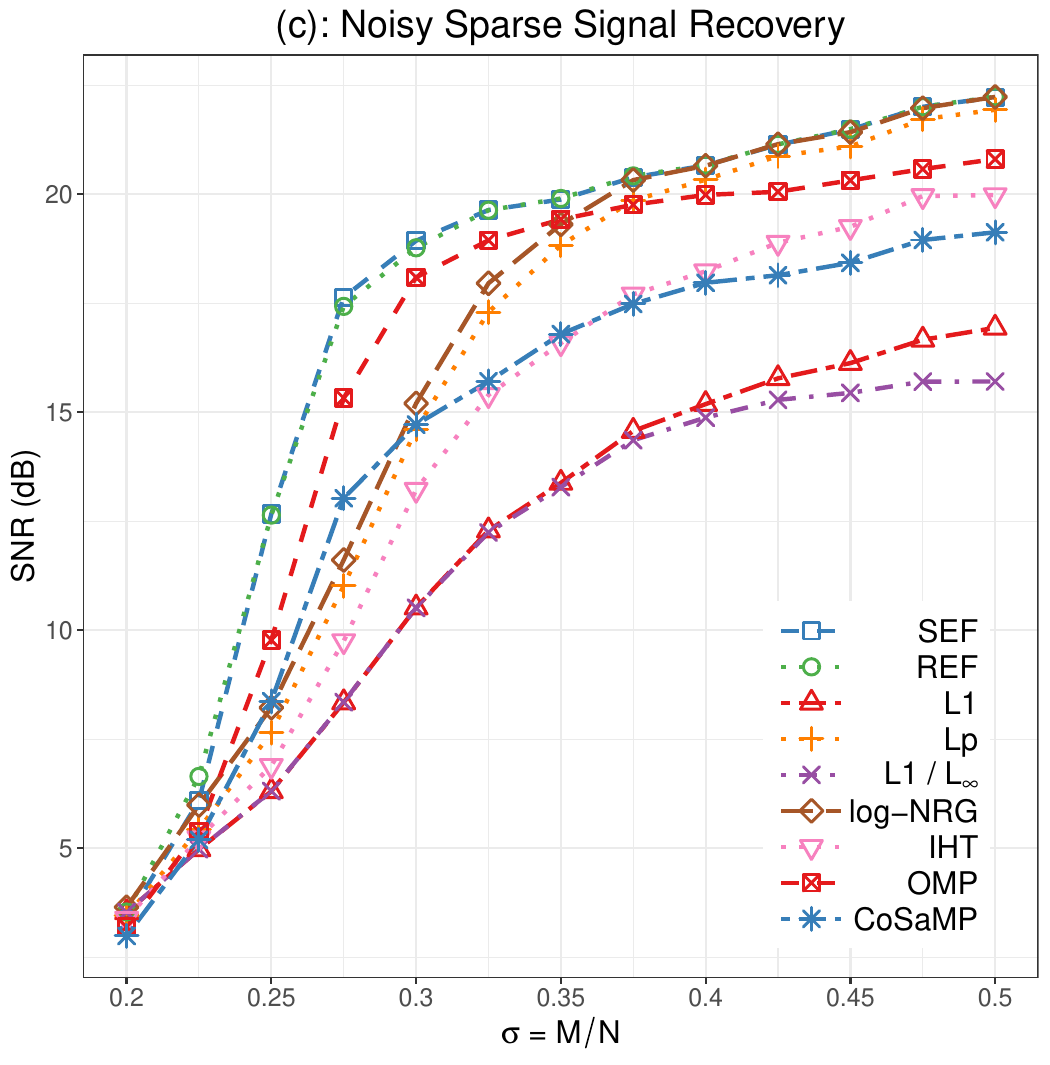}}
\vspace{-1mm}
\caption{\ref{fig:entropy_ptc_show} The phase transition curves of $h_p(\vx)$ and $h_{p,\alpha}(\vx)$ divide the $(\rho,\sigma)$ plane into the success phase and the failure phase; \ref{fig:ptc_noiseless_sparse} The phase transition curves of different sparsity regularization approaches in the noiseless case; \ref{fig:snr_noisy_sparse} The signal-to-noise-ratio (SNR) of the recovered signal $\vx$ using different sparsity regularization approaches in the noisy case.} 
\label{fig:ptc_show}
\end{figure*}

\section{Experimental Results}
We compare the proposed Shannon entropy function (SEF) and R\'{e}nyi entropy function (REF) regularizers with the relaxed sparsity-promoting regularizers introduced in section \ref{sec:sparsity_regularizers}, OMP \cite{OMP07}, CoSaMP \cite{CoSaMP09} and the iterative hard thresholding (IHT) approach \cite{Blumensath2008,Blumensath2009,Blumensath2010} that directly uses the $\|\vx\|_0$ as a regularizer on both simulated and real datasets. 

The accelerated inexact proximal gradient method in Section \ref{sec:acc_ipgm} can be adapted to solve \eqref{eq:sparsity_regularization} when $g(\vx)$ is some other lower semi-continuous nonconvex/nonsmooth regularizers. The first-order approximation \eqref{eq:approx_shannon_entropy} needs to be updated accordingly. For best performances and a fair comparison, the $l_1$-minimization solution obtained via FISTA is used as the initializer for all the other approaches.

Take the regularizers introduced in section \ref{sec:sparsity_regularizers} for example, the function $g_3(\vx)$ is a special case of $g_4(\vx)$ for signals with fixed $l_1$ norm; the functions $g_6(\vx)$, $\cdots$, $g_9(\vx)$ are special cases of the proposed entropy functions; the function $g_{10}(\vx)$ is not a sparsity-promoting regularizer. For the experiments, we thus compare the proposed entropy functions with the rest regularizers. 

Specifically, The $l_1$ norm $g_1(\vx)=\|\vx\|_1$ is minimized using FISTA. The $l_p$ norm $g_2(\vx)$ and the ``$\textnormal{L}_1/\textnormal{L}_\infty$'' function $g_4(\vx)$ are minimized using Algorithm \ref{alg:entropy_function_fista}, i.e. adapted FISTA. We show that $g_2(\vx)$, $g_4(\vx)$ are both lower semi-continuous and give their first-order approximations in Appendix \ref{app:other_regularizers}. The ``logarithm of energy'' $g_5(\vx)$ is not lower semi-continuous, its corresponding optimization problem is solved using the regularized FOCUSS algorithm \cite{RegFOCUSS03,FOCUSS97}: a reweighted $l_2$ norm minimization algorithm. 

The detailed experimental settings and code files can be found at \urlstyle{tt}\url{https://github.com/shuai-huang/GEFM}.

\begin{figure*}[tbp]
\centering
\subfigure{
\label{fig:psnr_noiseless_barbara_sparse}
\includegraphics[height=1.7in]{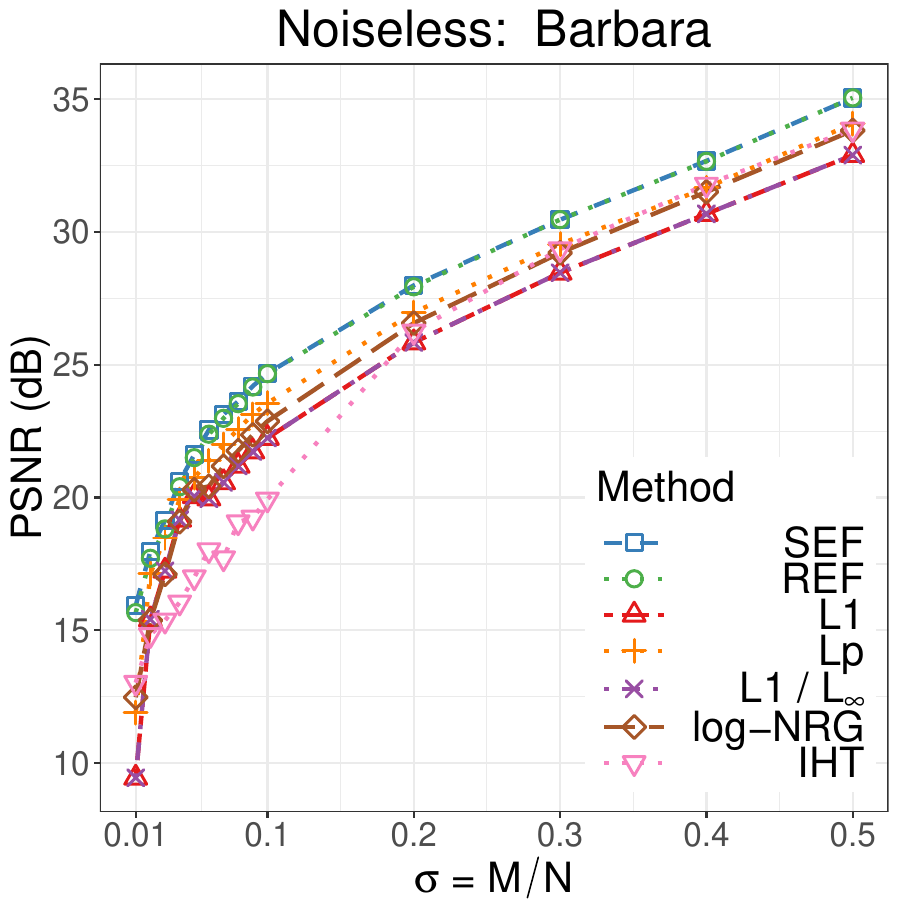}}
\subfigure{
\label{fig:psnr_noiseless_boat_sparse}
\includegraphics[height=1.7in]{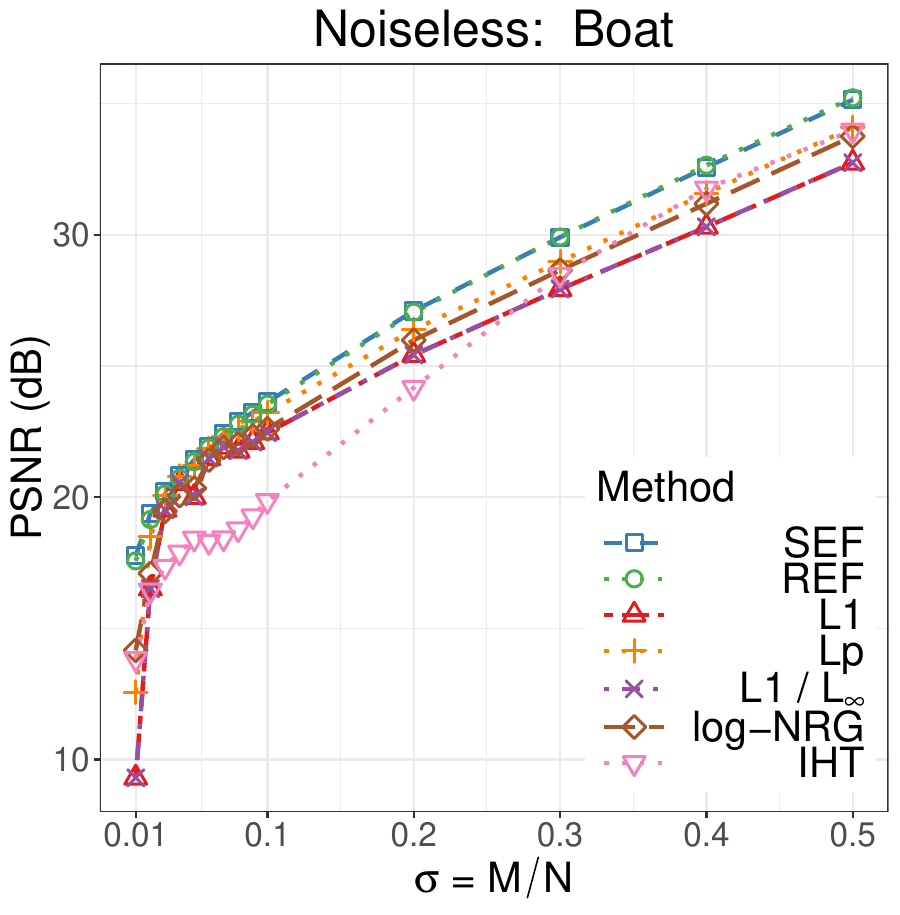}}
\subfigure{
\label{fig:psnr_noiseless_lena_sparse}
\includegraphics[height=1.7in]{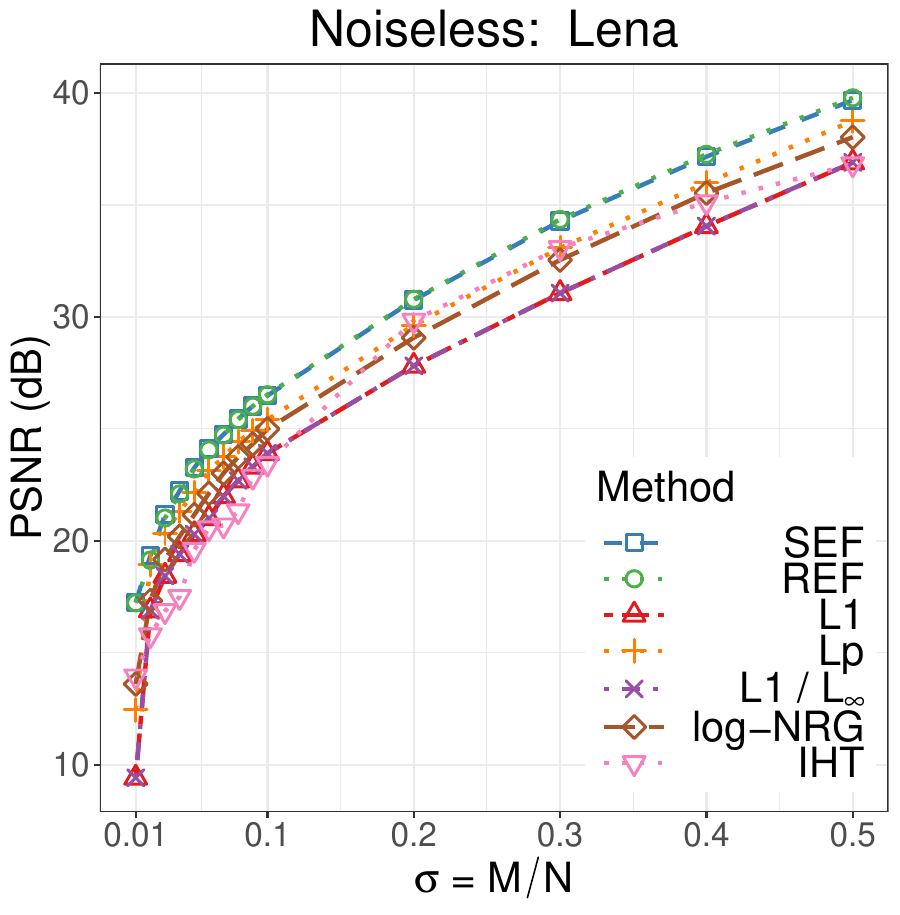}}
\subfigure{
\label{fig:psnr_noiseless_peppers_sparse}
\includegraphics[height=1.7in]{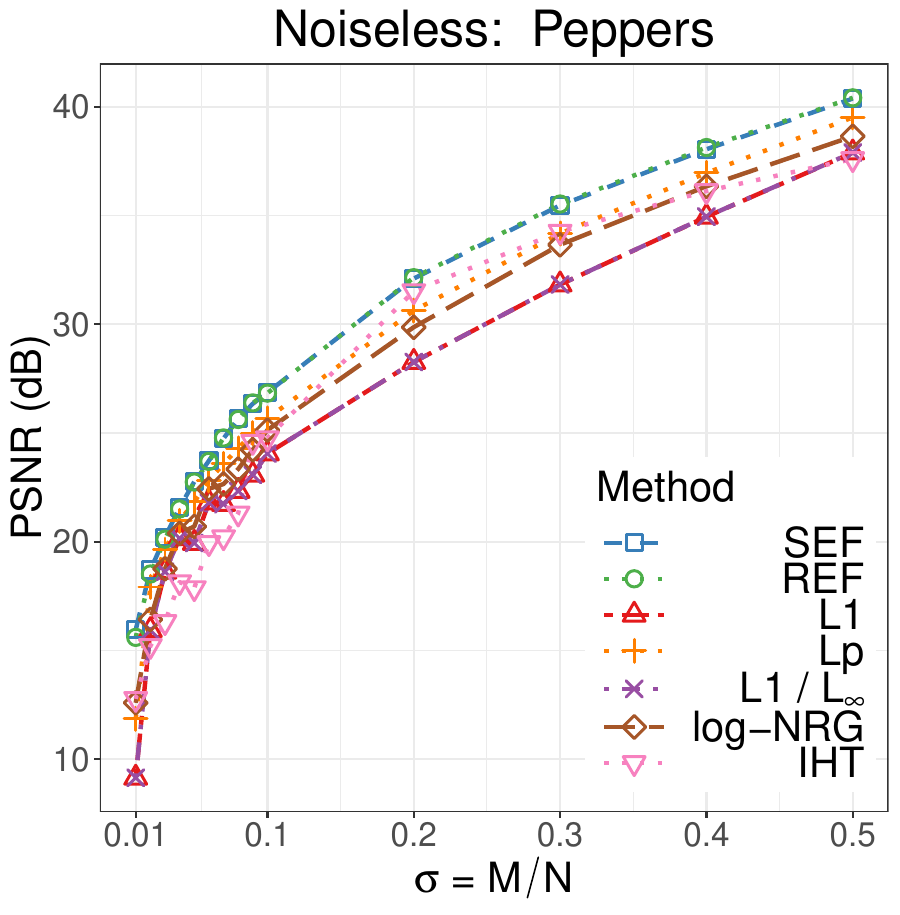}}\\
\vspace{-1mm}
\subfigure{
\label{fig:psnr_noisy_barbara_sparse}
\includegraphics[height=1.7in]{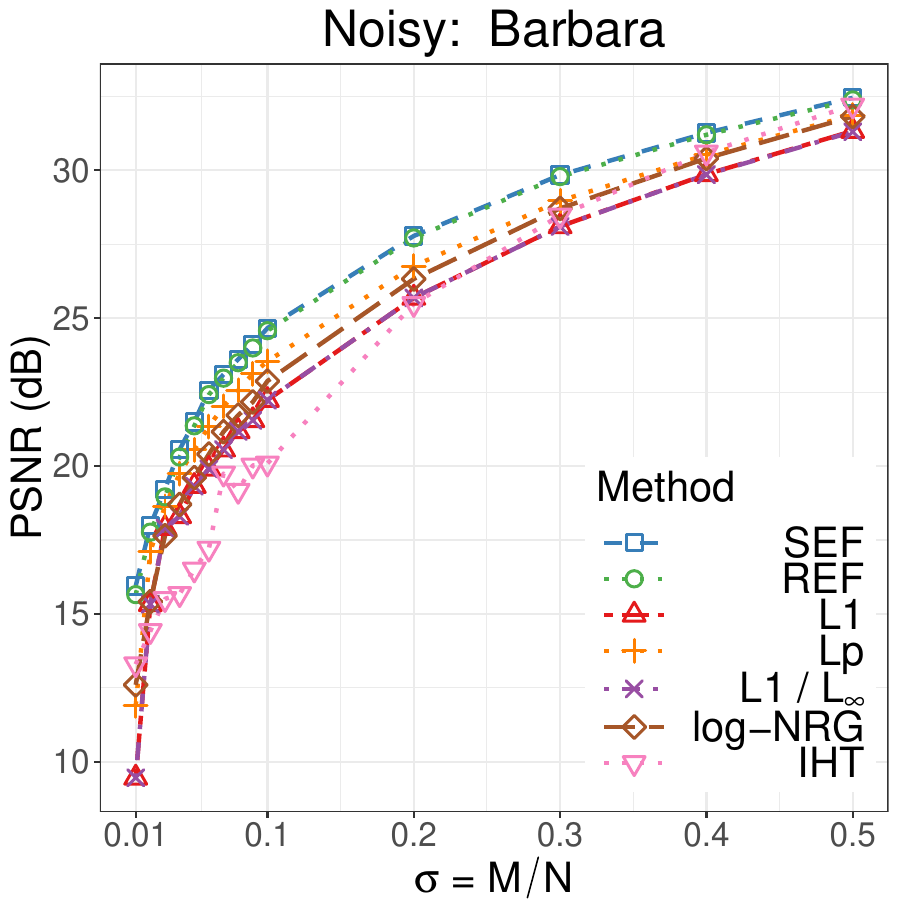}}
\subfigure{
\label{fig:psnr_noisy_boat_sparse}
\includegraphics[height=1.7in]{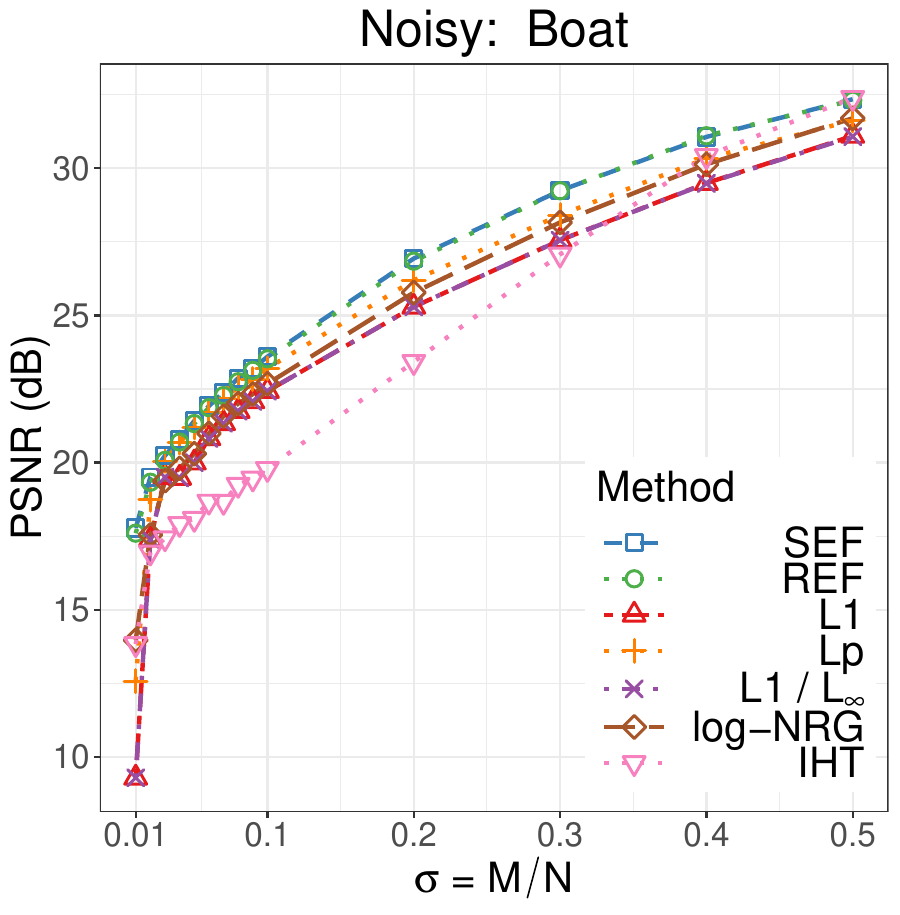}}
\subfigure{
\label{fig:psnr_noisy_lena_sparse}
\includegraphics[height=1.7in]{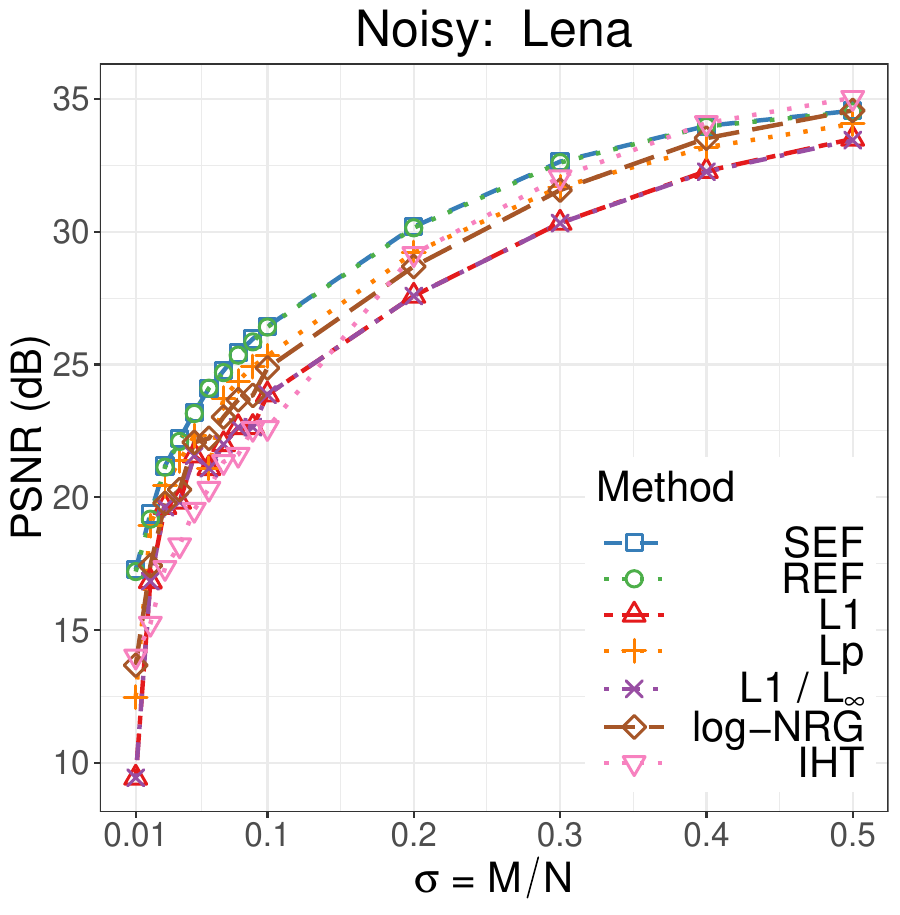}}
\subfigure{
\label{fig:psnr_noisy_peppers_sparse}
\includegraphics[height=1.7in]{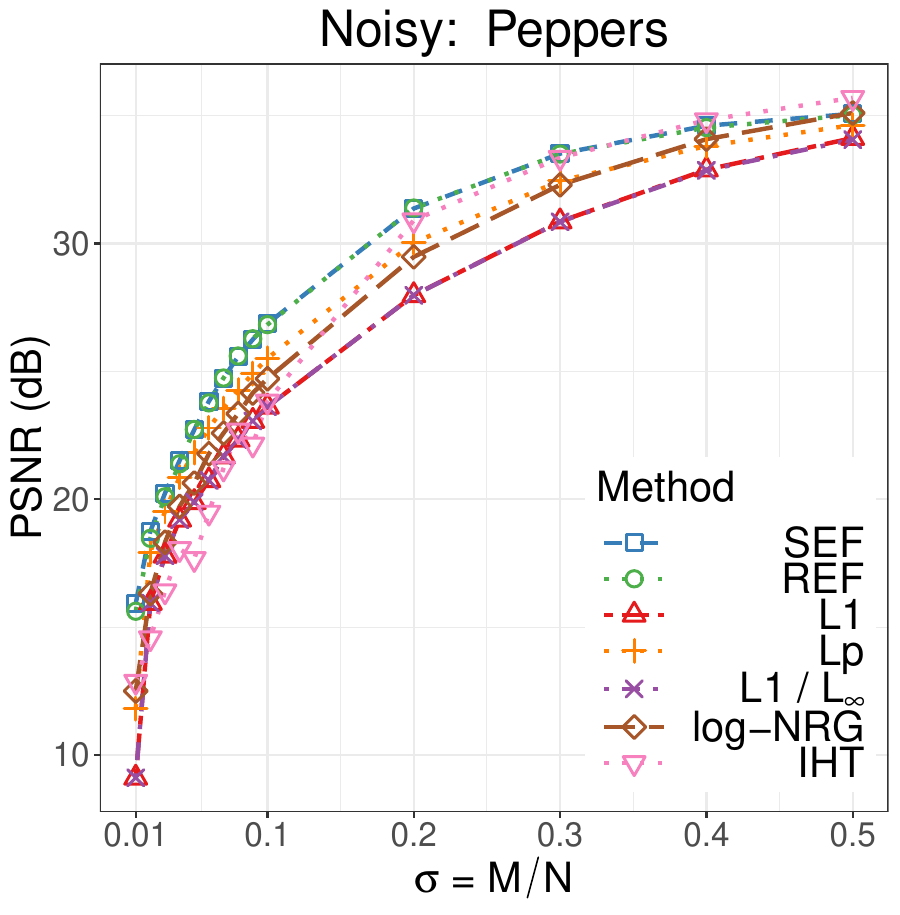}}
\vspace{-1mm}
\caption{The PSNR of the recovered images using different sparsity regularization approaches.} 
\label{fig:psnr_noiseless_noisy_images_sparse}
\end{figure*}

\subsection{Simulated Sparse Signal Recovery}
\subsubsection{Noiseless recovery}
For the noiseless sparse signal recovery experiments, we fix $N=1000$ and vary the sampling ratio $\sigma=\frac{M}{N}\in[0.05,0.1,\cdots,0.95]$ and the sparsity ratio $\rho=\frac{S}{M}\in[0.05,0.1,\cdots,0.95]$, where $S$ is the sparsity level of the signal, i.e. the number of nonzero coefficients. As mentioned in section \ref{sec:uniq_sol}, generally the sensing matrix $\vA$ needs to satisfy the RIP to guarantee the uniqueness of the recovered solution $\hat{\vx}$. For each combination of $\sigma$ and $\rho$, we randomly generate $100$ pairs of $\{\vx,\vA\}$: $\vA$ is a $M\times N$ random Gaussian matrix with normalized and centralized columns. The \emph{nonzero} entries of the sparse signal $\vx\in\mathbb{R}^N$ are i.i.d. generated according to the Gaussian distribution $\mathcal{N}(0,1)$.

Given the measurement vector $\vy=\vA\vx$ and the sensing matrix $\vA$, we try to recover the signal $\vx$. If $\epsilon=\|\hat{\vx}-\vx\|_2/\|\vx\|_2<10^{-3}$, the recovered $\hat{\vx}$ is considered to be a success. A decreasing $\lambda$ sequence was used to approach the optimal $\lambda=0$ as explained in section \ref{sec:regularization_para}. The parameters were tuned to obtain best performance for each method: for the SEF minimization approach, $p=1.1$; for the REF minimization approach, $p=1.1,\,\alpha=1.1$; for the Lp minimization approach, $p=0.5$. The $\textnormal{L}_1/\textnormal{L}_\infty$ function has a rigid energy-promoting behavior. When it is minimized, only the highest-energy coefficient(s) is/are promoted while the rest coefficients are suppressed. In our experiments, we found that the recovered $\hat{\vx}$ generally converged to its initializer, i.e. the $\textnormal{L}_1$-min solution. For the $\log$-energy minimization approach, as the lambda sequence went to $0$, the regularized FOCUSS algorithm was equivalent to the original FOCUSS algorithm \cite{FOCUSS97}. For the IHT approach, we scaled the sensing matrix $\vA$ by its operator norm to make $\|\vA\|_2<1$ as suggested by \cite{Blumensath2008} to make the algorithm converge. For the OMP and CoSaMP, the sparsity level $S$ is tuned to obtain best performances. The measurement $\vy$ was also scaled accordingly. Based on the $100$ trials, we computed the success recovery rate for each combination of $\{\sigma,\,\rho\}$, and plotted the phase transition curves (PTC) in Fig. \ref{fig:ptc_noiseless_sparse}. 

As shown in Fig. \ref{fig:entropy_ptc_show}, the PTC is the contour that corresponds to the 0.5 success rate in the domain $\{\sigma,\rho\}\in\{0,1\}^2$, it divides the domain into a ``success'' phase (lower right) and a ``failure'' phase (upper left). We can see that the proposed SEF minimization and REF minimization approaches generally perform equally well, and they both perform better than the other approaches. 

\subsubsection{Noisy recovery}
We next try to recover the sparse signal $\vx$ from a noisy measurement vector $\vy$. Specifically, we fix $S=100,N=1000$ and increase $M$ gradually. The measurement $\vy\in\mathbb{R}^M$ is generated as follows:
\begin{align}
\textrm{Noisy measurements:}\quad\vy=\vA\vx+\nu\cdot\vw\,,
\end{align}
where $\nu>0$ controls the amount of noise added to $\vy$, the entries of $\vw$ are i.i.d Gaussian $\mathcal{N}(0,1)$. We choose $\nu=0.05$, this creates a measurement $\vy$ with signal to noise ratio (SNR) around $25$ dB. We randomly generate $100$ triples of $\{\vx,\vA,\vw\}$. The average SNRs of the recovered signals $\hat{\vx}$ are shown in Fig. \ref{fig:snr_noisy_sparse}. We can see that when $\sigma<0.4$, the SEF and REF minimization approaches outperform the other approaches.

\subsection{Real Image Recovery}
\begin{figure}[tbp]
\centering
\subfigure{
\includegraphics[width=0.75in]{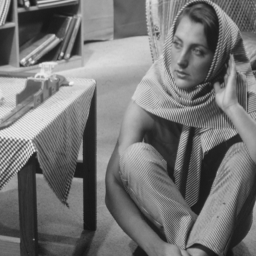}}
\subfigure{
\includegraphics[width=0.75in]{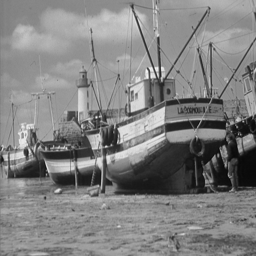}}
\subfigure{
\includegraphics[width=0.75in]{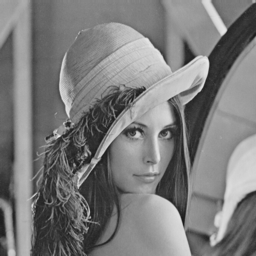}}
\subfigure{
\includegraphics[width=0.75in]{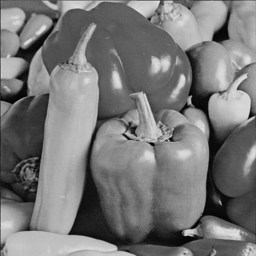}}
\caption{The real images used in the recovery experiments (from left to right): Barbara, Boat, Lena, Peppers.} 
\label{fig:real_images}
\end{figure}

Real images are considered to be approximately sparse under some proper basis, such as the DCT basis, wavelet basis, etc. Here we compare the recovery performances of the aforementioned sparsity regularization approaches based on varying noiseless and noisy measurements of the $4$ real images shown in Fig. \ref{fig:real_images}: Barbara, Boat, Lena, Peppers. Specifically, in order to reveal the sparse coefficients $\vx$ of the real images $\vs$, we use the sparsity averaging method by \cite{SACS13} to construct an over-complete wavelet basis by concatenating Db1-Db4 \cite{DBWav92}:
\begin{align}
\label{eq:wavelet_concat}
\vs=\Psi(\vx) = \frac{1}{2}\sum_{i=1}^4\Psi_{\textnormal{Db}i}(\vx)\,,
\end{align}
where $\Psi_{\textnormal{Db}i}$ is the corresponding operator of the Db$i$ wavelet. There is little performance improvement when adding more than $4$ wavelet operators. The $\frac{1}{2}$ is introduced in \eqref{eq:wavelet_concat} to ensure $\vs = \Psi(\vx)$, and $\vx =\Psi^\dag(\vs)$.

The sampling matrix $\vU$ is constructed using the structurally random matrix approach by \cite{SRM12}: $\vU=\vD\vF\vR$, where $\vR$ is a uniform random permutation matrix that scrambles the signal's sample locations globally while a diagonal matrix of Bernoulli random variables flips the signal's sample signs locally, $\vF$ is an orthonormal DCT matrix that computes fast transforms, $\vD$ is a sub-sampling matrix that randomly selects a subset of the rows of the matrix $\vF\vR$. 

The noiseless measurements $\vy$ of the image $\vs$ are obtained:
\begin{align}
\vy = \vU\vs\,.
\end{align}
The noisy measurements $\vy$ are obtained:
\begin{align}
\vy = \vU\vs+\nu\cdot\vw\,.
\end{align}
The entries of the noise $\vw$ are generated using i.i.d. Gaussian distribution $\mathcal{N}(0,1)$, $\nu$ is chosen to be $0.02$ so that the SNR of the measurement vector $\vy$ is around $30$ dB\footnote{When $\nu$ is set to other values, the relative performances of the four methods are similar.}. 

Take the SEF minimization for example, the corresponding recovery problem is a cosparse problem \cite{NAM201330} with respect to the image $\vs$:
\begin{align}
\min_{\vs}\,\|\vy-\vU\vs\|_2^2+\lambda\, h_p\left(\Psi^\dag(\vs)\right)\,.
\end{align}
The inexact proximal operator in the $(t+1)$-th iteration is:
\begin{align}
\label{eq:real_image_s1}
\begin{split}
&\textnormal{prox}_{\frac{\lambda}{\kappa}h_p}\left(\vs^{(t)}-\frac{2}{\kappa}\cdot\vU^\textrm{T}(\vU\vs^{(t)}-\vy)\right)\\
&=\arg\min_{\vs}\,\frac{\kappa}{2}\left\|\vs-\left(\vs^{(t)}-\frac{2}{\kappa}\cdot\vU^\textrm{T}(\vU\vs^{(t)}-\vy)\right)\right\|_2^2\\
&\quad\quad\quad\quad\quad+\lambda\,h_p\left(\Psi^\dag(\vs)\right),
\end{split}
\end{align}
where $\kappa=2$ for the chosen sampling matrix $\vU$. The inexact proximal operator (\ref{eq:real_image_s1}) can be efficiently solved using the alternating split bregman shrinkage algorithm by \cite{Setzer2009}.

Since the real images are only approximately sparse, both the noiseless and noisy recovery experiments are performed using a fixed $\lambda$. The parameters were tuned to obtain best performance for each approach\footnote{Since no direct matrix-vector product form is available for \ref{eq:wavelet_concat}, OMP and CoSaMP are not compared in the image recovery experiments.}. For the SEF minimization approach, $p=1$, $\lambda=5000$; for the REF minimization approach, $p=0.9$, $\alpha=1.1$, $\lambda=10^4$; for the $\textnormal{L}_1$ minimization approach, $\lambda=0.1$; for the Lp minimization approach, $p=0.8$, $\lambda=10^{-2}$; for the $\textnormal{L}_1/\textnormal{L}_\infty$ minimization approach, $\lambda=\frac{1}{10}N$; for the $\log$-energy minimization approach, $\lambda=1$; for the IHT approach, $\lambda=5$. The peak signal to noise ratios (PSNR) of the noiseless and noisy recovery experiments are shown in Fig. \ref{fig:psnr_noiseless_noisy_images_sparse}. We can see that the proposed SEF and REF minimization approaches perform equally well, and they generally have the best performances in terms of PSNR (dB). In the noisy recovery of ``Lena'' and ``Peppers'', IHT performs slightly better than the entropy functions minimizations when $\sigma\geq 0.4$.

\subsection{Face Recognition via Sparse Representation Classification}
As introduced earlier in Section \ref{sec:adv_src}, SRC was proposed in \cite{SRC_face09,SRC_CVPR10} as an effective approach to perform the classification task based on the sparse representation of a test sample $\vy$ in the linear subspace spanned by the labeled training samples $\vA$. The sparse representation $\vx$ is computed by solving the following problem \cite{SRC_face09}:
\begin{align}
\arg\min_{\vx}\quad\quad\|\vy-\vA\vx\|_2^2+\lambda\, g(\vx)\,, \tag{\ref{eq:sparsity_regularization} revisited}
\end{align}
where $g(\vx)$ is some sparsity-promoting regularizer. \cite{SRC_face09} used the $l_1$-norm $\|\vx\|_1$ as the regularizer and tested the SRC model on the face recognition task. Significant and robust improvements over other popular approaches were observed across multiple benchmark datasets .

\begin{figure}[tbp]
\centering
\subfigure{
\includegraphics[width=0.9in]{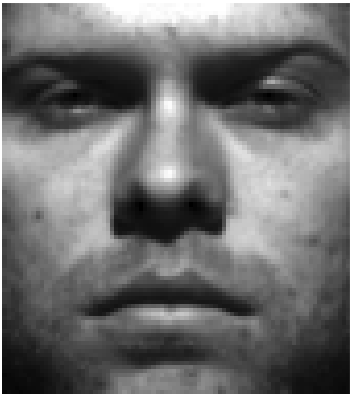}}
\subfigure{
\includegraphics[width=0.9in]{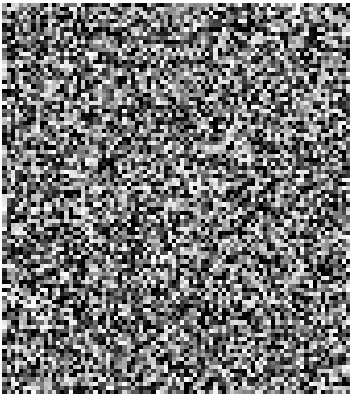}}
\subfigure{
\includegraphics[width=0.9in]{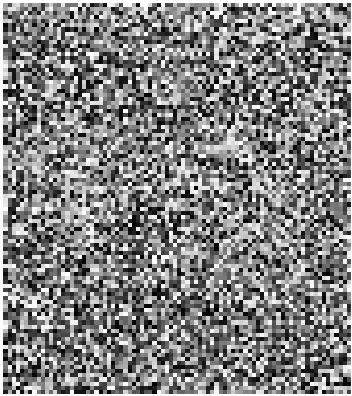}}
\caption{Random pixel corruption of the face image \cite{SRC_face09} (from left to right): The noiseless image; The i.i.d random noise; The image with $90\%$ of the pixels corrupted.} 
\label{fig:random_corruption}
\vspace{-2mm}
\end{figure}

Following the same experimental setup as \cite{SRC_face09}, we choose the challenging task of face recognition under random pixel corruption and compare the performances of the aforementioned sparsity regularization approaches. Specifically, each image in the Extended Yale B Face Database \cite{YaleFaceB01} is resized to $96\times84$ pixels and then converted to a vector via column-concatenation. Each test image is randomly corrupted by replacing the pixel value with a number uniformly chosen from [0, 255], as show in Fig. \ref{fig:random_corruption}. 

\begin{figure}[tbp]
\centering
\includegraphics[height=2.8in]{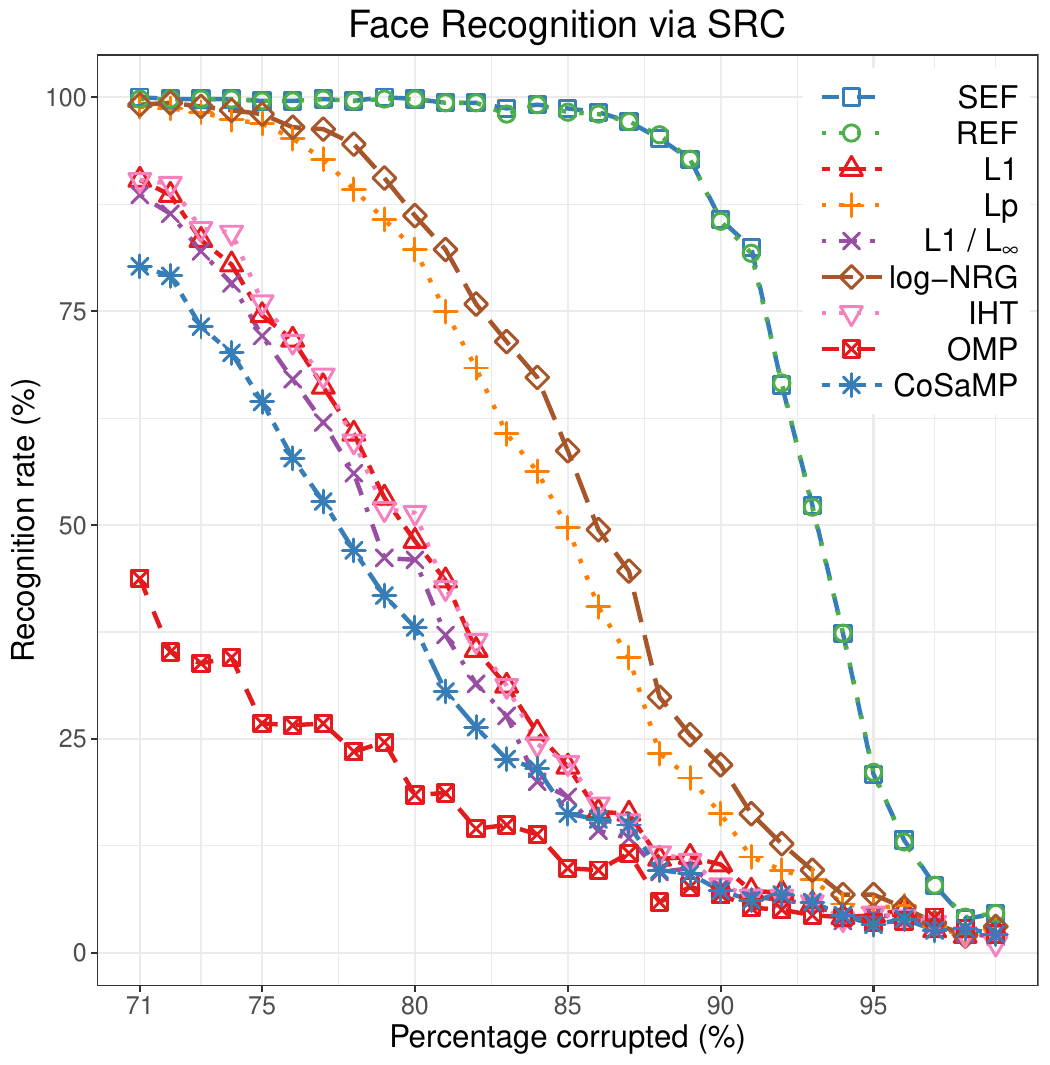}
\caption{Face recognition based on SRC under random pixel corruption using different sparsity regularization approaches.} 
\label{fig:src_compare}
\vspace{-2mm}
\end{figure}

In this case, the dictionary $\vA$ is composed of two sub-dictionaries: $\vA=[\vA_{tr}\quad\vI]$. $\vA_{tr}$ is the collection of noiseless training samples that contributes to the ``uncorrupted'' test image, while $\vI$ is the identity matrix that contributes to the random noise. $\vx$ can then be separated into two parts accordingly: $\vx^T=[\vx_{tr}^T\quad\vx_{\vI}^T]$. Based on the observation that the respective distributions of the entries in $\vx_{tr}$ and $\vx_{\vI}^T$ are quite different, we propose to regularize them using different regularization parameters as follows:
\begin{align}
\arg\min_{\vx}\quad\quad\|\vy-\vA\vx\|_2^2+\lambda\cdot g(\vx_{tr})+\lambda\mu\cdot g(\vx_{\vI})\,.
\end{align}
As done before, the parameters were tuned for each approach separately to obtain best performances. For the SEF minimization approach, $p=1.2$, $\lambda=10^{-3}$, $\mu=75$; for the REF minimization approach, $p=1.15$, $\alpha=1.1$, $\lambda=10^{-3}$, $\mu=75$; for the $\textnormal{L}_1$ minimization approach, $\lambda=10^{-3}$, $\mu=2.5$; for the Lp minimization approach, $p=0.5$, $\lambda=10^{-4}$, $\mu=5$; for the $\textnormal{L}_1/\textnormal{L}_\infty$ minimization approach, $\lambda=10^{-6}$, $\lambda=2.5$; for the $\log$-energy minimization approach, $\lambda=10^{-6}$, $\mu=5$; for the IHT approach, $\lambda=10^{-3}$, $\mu=6$. For the OMP and CoSaMP, the sparsity constraint on the training samples is set to $S_{tr}\leq 20$, corresponding to the average number of samples per class in the experiments; the sparsity constraint on the noise is set to $S_{\vI}\leq 7000$. The recognition rates of various approaches under increasing percentages of random pixel corruptions are shown in Fig. \ref{fig:src_compare}. We can see that the sparse representations obtained by minimizing the proposed SEF and REF lead to the highest recognition rates. The performance only starts to drop significantly when more than $90\%$ of the pixels are corrupted.

\section{Conclusion and Future Work}
\label{sec:conclusion}
In this paper we proposed the \emph{generalized} Shannon entropy function $h_p(\vx)$ and R\'{e}nyi entropy function $h_{p,\alpha}(\vx)$ as the sparsity regularizers for the sparse signal recovery task. Regardless of the values of $p\in(0,\infty]$ and $\alpha\in(0,1)\cup(1, \infty]$, the local minimums of the generalized entropy functions occur on the boundaries of the orthants in $\mathbb{R}^N$, and minimizing them produces sparse solutions. Both $h_p(\vx)$ and $h_{p,\alpha}(\vx)$ are nonconvex functions, the corresponding minimization problems \eqref{eq:sefm_ssr},\eqref{eq:refm_ssr} can be solved via the adapted FISTA until convergence. Compared to other sparsity regularizers, minimizing the generalized entropy functions adaptively promotes multiple high-energy coefficients while suppressing the rest low-energy coefficients of the recovered solution. 

Sparse signal recovery experiments on both the simulated and real data show the proposed entropy function regularizations perform better than OMP, CoSaMP and other popular sparsity regularizers, and achieve state-of-the-art performances. This motivates us to explore sufficient conditions for the uniqueness of the solution, seek matrices that satisfy such uniqueness conditions, and establish error bounds on the recovered signal $\hat{\vx}$ in our future work. Additionally, we would like to apply the generalized entropy functions minimization approaches to other Compressive Sensing applications such as RPCA \cite{RPCA_nips09, lin2009augmented, Lin09fastconvex, RPCA_candes11}, dictionary learning\cite{KSVD06,IMDenoise06,OLDC09}, etc.

\begin{appendices}
\counterwithin{assumption}{section}
\counterwithin{theorem}{section}
\renewcommand\thetable{\thesection\arabic{table}}
\renewcommand\thefigure{\thesection\arabic{figure}}

\section{Monotonic Objective Function Minimization}
\label{app:mono_min}
Take the Shannon entropy function minimization (SEF) problem $P_{h_p}(\vx)$ in (\ref{eq:sefm_ssr}) for example, we have the following:
\begin{align}
\label{eq:decrease_fun}
\begin{split}
F(\vr)&=f(\vr)+\lambda h_p(\vr)\\
&\stackrel{\textnormal{(a)}}{\leq} f(\vx^{(t)})+\left<\vr-\vx^{(t)}, \nabla f(\vx^{(t)})\right>\\
&\quad+\frac{\kappa}{2}\left\|\vr-\vx^{(t)}\right\|_2^2+\lambda h_p(\vr)\\
&=f(\vx^{(t)})-\frac{1}{2\kappa}\|\nabla f(\vx^{(t)})\|_2^2\\
&\quad+\frac{\kappa}{2}\left\|\vr-\left(\vx^{(t)}-\frac{1}{\kappa}\nabla f(\vx^{(t)})\right)\right\|_2^2+\lambda h_p(\vr)\\
&\stackrel{\textnormal{(b)}}{\leq}f(\vx^{(t)})-\frac{1}{2\kappa}\|\nabla f(\vx^{(t)})\|_2^2\\
&\quad+\frac{\kappa}{2}\left\|\vx^{(t)}-\left(\vx^{(t)}-\frac{1}{\kappa}\nabla f(\vx^{(t)})\right)\right\|_2^2+\lambda h_p(\vx^{(t)})\\
&=f(\vx^{(t)})+\lambda h_p(\vx^{(t)})\\
&=F(\vx^{(t)})\,.
\end{split}
\end{align}
The first inequality ``(a)'' is obtained using (\ref{eq:proximal_data_fidelity}). The second inequality ``(b)'' holds because we enforce \eqref{eq:min_constraint_r} when the nonconvex proximal operator \eqref{eq:proximal_shannon_en_fun} is inexactly solved. Combined with \eqref{eq:update_acceleration_x}, the proposed Algorithm \ref{alg:entropy_function_fista} is monotonic.

Additionally, we have:
\begin{itemize}
\item $h_p(\vx)$ is continuous in $\mathbb{R}^N$, hence it is also lower semi-continuous. 
\item The data fidelity term $f(\vx)=\|\vy-\vA\vx\|_2^2\geq 0$ is bounded from below. $f(\vx)\rightarrow\infty$ when $\|\vx\|_2\rightarrow\infty$, hence $f(\vx)$ is coercive. $h_p(\vx)\geq 0$ is also bounded from below. Hence $F(\vx)=f(\vx)+\lambda h_p(\vx)$ is coercive.
\item $f(\vx)$ has Lipschitz-continuous gradients \cite{FISTA09}.
\end{itemize}  
Using Theorem $1$ of \cite{APG2015}, we can get that the proposed Algorithm \ref{alg:entropy_function_fista} converges to a critical point.

\begin{figure*}[tbp]
\centering
\subfigure{
\label{fig:gsst_1}
\includegraphics[height=1.5in]{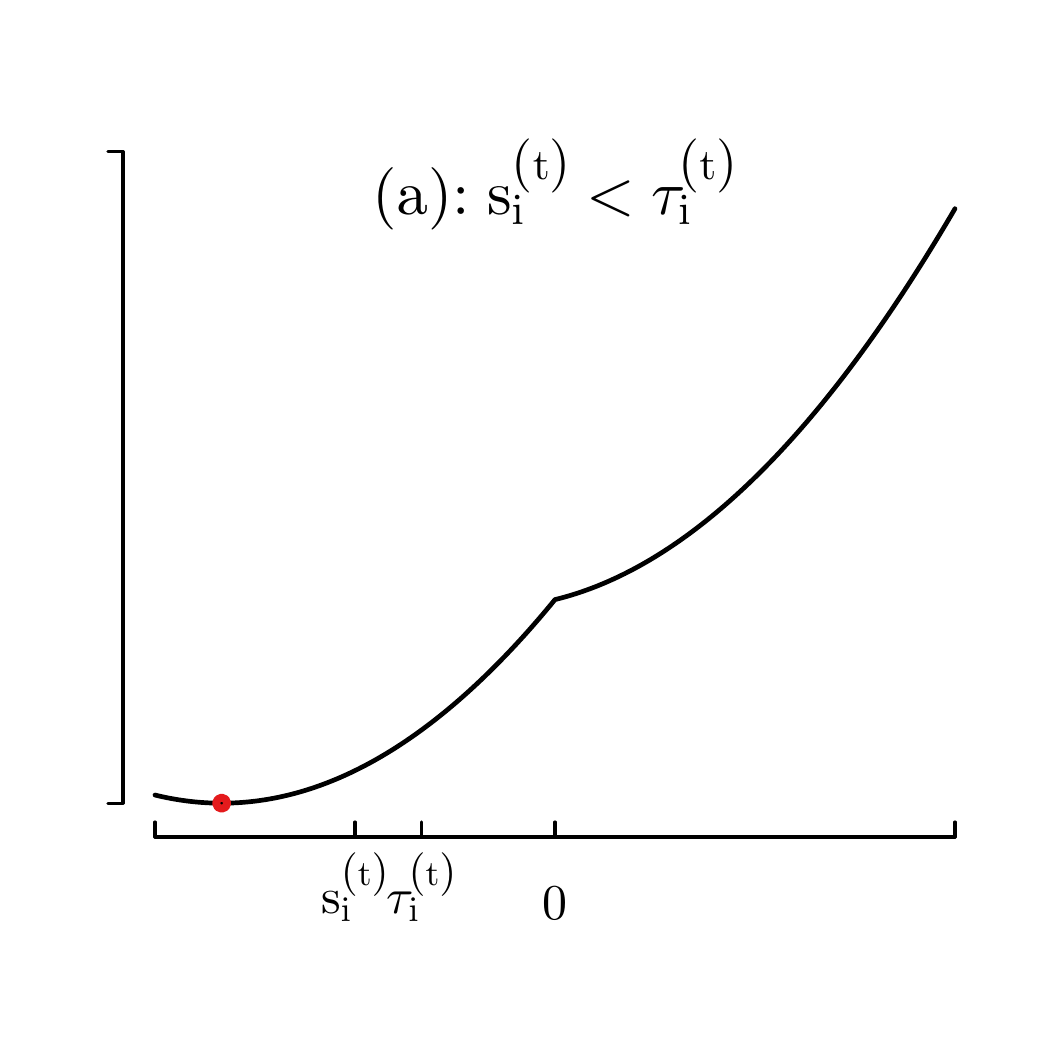}}
\subfigure{
\label{fig:gsst_2}
\includegraphics[height=1.5in]{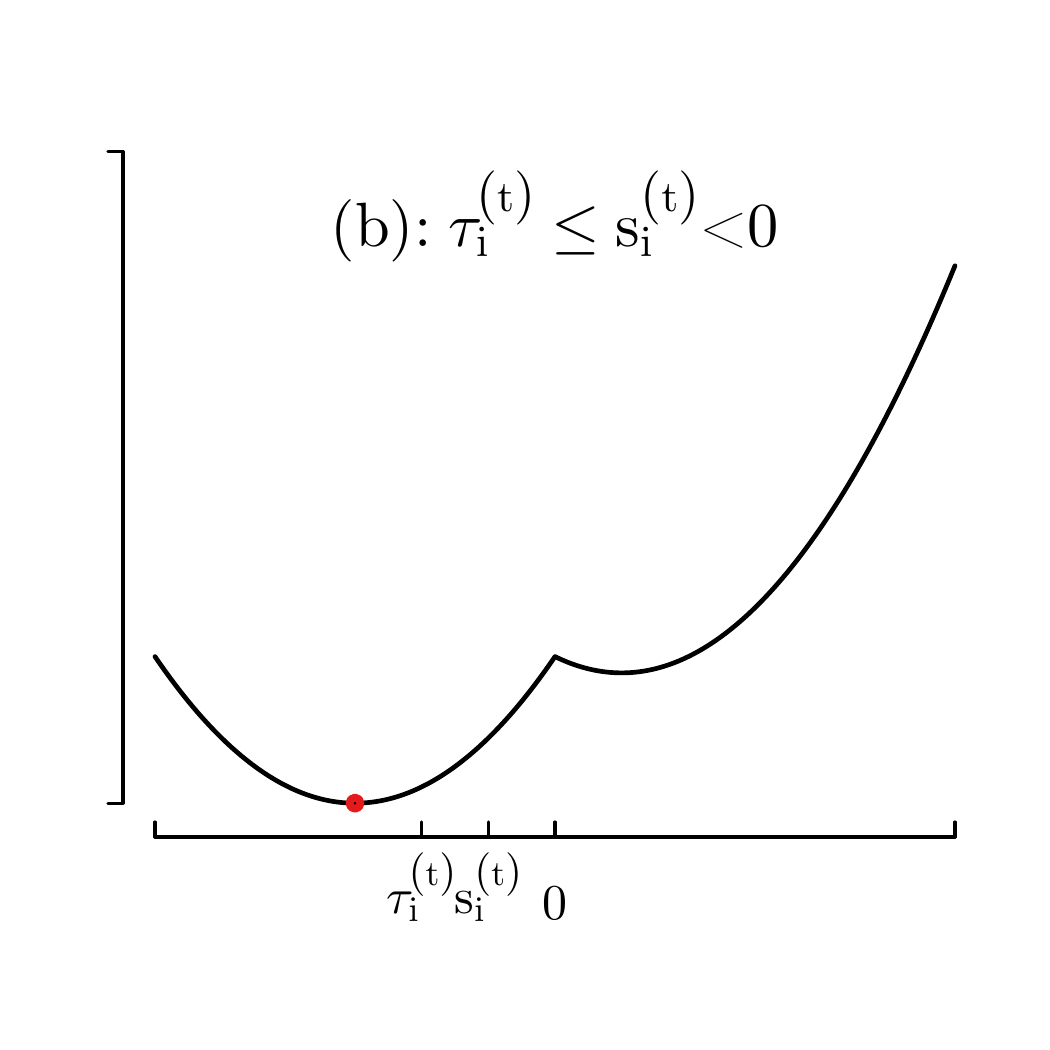}}
\subfigure{
\label{fig:gsst_3}
\includegraphics[height=1.5in]{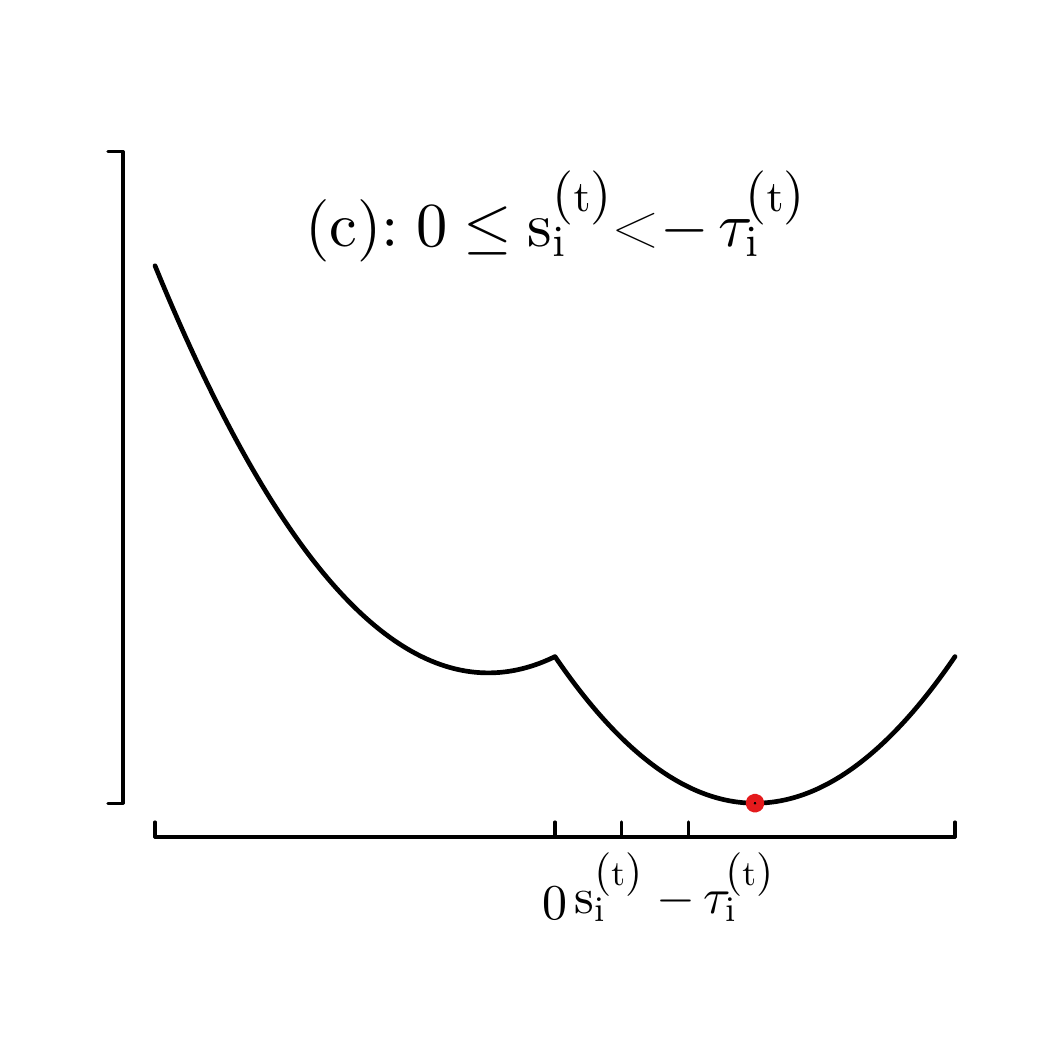}}
\subfigure{
\label{fig:gsst_4}
\includegraphics[height=1.5in]{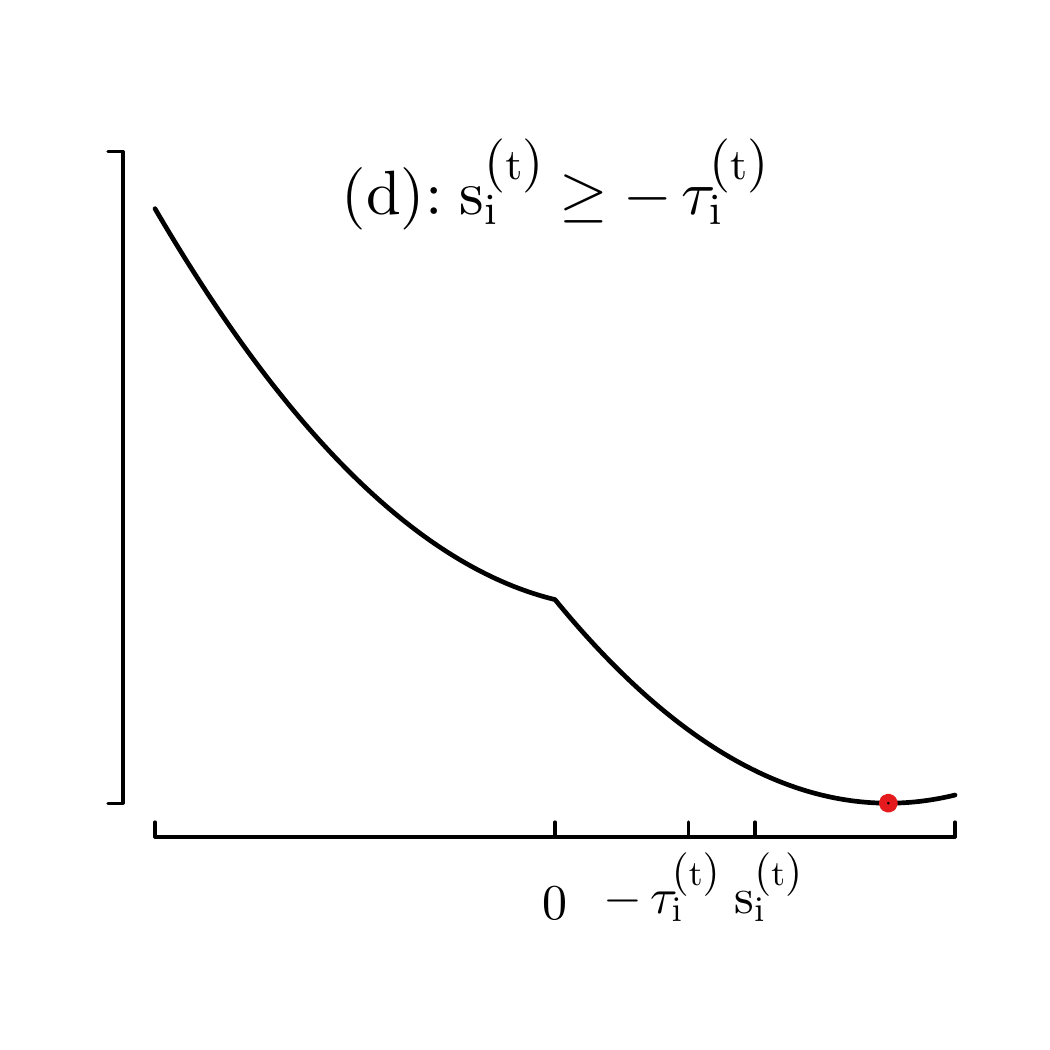}}
\caption{Nonconvex soft shrinkage threshoding when the threshold $\tau^{(t)}<0$.} 
\label{fig:gsst}
\end{figure*}

\section{Energy-promoting Analysis of other regularization functions}
\label{app:energy_promoting_other}
Here we discuss the energy-promoting behaviors of the other sparsity regularizers\footnote{$g_{10}(\vx)$ is not a sparsity regularizer.} introduced in section \ref{sec:sparsity_regularizers} from an optimization perspective. Specifically, the regularizers are minimized via subderivative descent:
\begin{align}
\vx &= \vx-\eta\cdot\partial h_p(\vx)  \tag{\ref{eq:gd_update_shannon} revisited}\,.
\end{align}
\begin{itemize}
\item The subderivatives of $g_1(\vx)$, $g_2(\vx)$, $g_5(\vx)$ are as follows:
\begin{align}
\frac{\partial }{\partial x_i}\|\vx\|_1&=\textnormal{sign}(x_i)\\
\frac{\partial }{\partial x_i}\|\vx\|_p^p&=\textnormal{sign}(x_i)\cdot p|x_i|^{p-1}\\
\frac{\partial }{\partial x_i}\sum_{i:|x_i|\neq 0}\log |x_i|^2 &= \textnormal{sign}(x_i)\cdot\frac{2}{|x_i|}\,.
\end{align}
The update \eqref{eq:gd_update_shannon} would decrease the magnitude of every coefficient. 

\item The subderivatives of $g_3(\vx)$, $g_4(\vx)$ are:
\begin{align}
\frac{\partial }{\partial x_i}\frac{1}{\|\vx\|_\infty}&=\textnormal{sign}(x_i)\cdot\frac{-\delta_\infty(i)}{\|\vx\|_\infty^2}\\
\frac{\partial }{\partial x_i}\frac{\|\vx\|_1}{N\|\vx\|_\infty} &= \frac{\textnormal{sign}(x_i)}{N}\cdot\left(\frac{1}{\|\vx\|_\infty}-\delta_\infty(i)\frac{\|\vx\|_1}{\|\vx\|_\infty^2}\right)\,,
\end{align}
where $\delta_\infty(i)$ is defined as follows:
\begin{align}
\label{eq:delta_infty}
&\delta_\infty(i)=\left\{
\begin{array}{l}
0\\
1
\end{array}
\quad
\begin{array}{l}
\textnormal{if $|x_i|\neq \max_j|x_j|$}\\
\textnormal{if $|x_i|=\max_j|x_j|$}\,.
\end{array}
\right.
\end{align}
The update \eqref{eq:gd_update_shannon} would only increase the magnitude(s) of the \emph{highest}-energy coefficient(s) for $g_3(\vx)$ and $g_4(\vx)$, while decreasing the magnitudes of the rest coefficients for $g_4(\vx)$. 

We should note that the energy-promoting behaviors of $g_3(\vx)$ and $g_4(\vx)$ are different from those of the proposed entropy functions. For the entropy functions, the thresholds $\nu$ in \eqref{eq:pd_nu_sef}, \eqref{eq:pd_nu_ref} are adaptive in that they are updated every iteration along with the solution, as opposed to the rigid criterion induced by $g_3(\vx)$ and $g_4(\vx)$.
\end{itemize}

\section{Nonconvex Soft Shrinkage Thresholding}
\label{app:gsst}

Take the optimization problem in (\ref{eq:shannon_en_approx_secondstep}) for example, let $\tau_i^{(t)}=\frac{\lambda}{\kappa}\nabla h_p(|x_i^{(t)}|)$, $ s_i^{(t)}=x_i^{(t)}-\frac{1}{\kappa}\nabla f(x_i^{(t)})$, we have the following problem for the $i$-th entry $x_i$:
\begin{align}
\label{eq:1d_shannon}
\min_{x_i}\quad\frac{1}{2}\|x_i- s_i^{(t)}\|_2^2+\tau_i^{(t)}|x_i|\,.
\end{align}

\begin{positive_threshold}
When $\tau_i^{(t)}\geq0$, (\ref{eq:1d_shannon}) is a convex problem. Its solution is given by applying the shrinkage operator given in (\ref{eq:shrinkage}) on $ s_i^{(t)}$ with the threshold $\tau_i^{(t)}$.
\end{positive_threshold}

\begin{negative_threshold}
When $\tau_i^{(t)}<0$, (\ref{eq:1d_shannon}) is a not necessarily a convex problem. Luckily this is a simple one dimensional problem, its global optimal solution can be still found as follows:
\begin{enumerate}
\item When $ s_i^{(t)}<\tau_i^{(t)}$, as shown in Fig. \ref{fig:gsst_1}:
\begin{itemize}
\item For $x_i\geq0$, we have:
\begin{align}
\label{eq:1d_shannon_1}
\begin{split}
\min_{x_i}\quad&\left(x_i+\tau_i^{(t)}- s_i^{(t)}\right)^2\\
&-\left(\tau_i^{(t)}- s_i^{(t)}\right)^2+\left( s_i^{(t)}\right)^2\,.
\end{split}
\end{align}
Since $\tau_i^{(t)}- s_i^{(t)}>0$, the $x_i$ that minimizes (\ref{eq:1d_shannon_1}) is $0$. 

\item For $x_i<0$, we have:
\begin{align}
\label{eq:1d_shannon_2}
\begin{split}
\min_{x_i}\quad&\left(x_i-\tau_i^{(t)}- s_i^{(t)}\right)^2\\
&-\left(\tau_i^{(t)}+ s_i^{(t)}\right)^2+\left( s_i^{(t)}\right)^2\,.
\end{split}
\end{align}
Since $-\tau_i^{(t)}- s_i^{(t)}>0$, the $x_i$ that minimizes (\ref{eq:1d_shannon_2}) is $\tau_i^{(t)}+ s_i^{(t)}$.
\end{itemize}

(\ref{eq:1d_shannon}) is continuous at the point $x_i=0$. Hence the global minimum of (\ref{eq:1d_shannon}) is obtained by $x_i=\tau_i^{(t)}+ s_i^{(t)}$.

\item When $\tau_i^{(t)}\leq s_i^{(t)}<0$, as shown in Fig. \ref{fig:gsst_2}:
\begin{itemize}
\item For $x_i\geq0$, we have (\ref{eq:1d_shannon_1}). Since $\tau_i^{(t)}- s_i^{(t)}<0$, the $x_i$ that minimizes (\ref{eq:1d_shannon_1}) is $-\tau_i^{(t)}+ s_i^{(t)}$. 

\item For $x_i<0$, we have (\ref{eq:1d_shannon_2}). Since $-\tau_i^{(t)}- s_i^{(t)}>0$, the $x_i$ that minimizes (\ref{eq:1d_shannon_2}) is $\tau_i^{(t)}+ s_i^{(t)}$.
\end{itemize}
It's easy to verify that the minimum of (\ref{eq:1d_shannon_2}) is \emph{smaller} than the minimum of (\ref{eq:1d_shannon_1}). Hence the global minimum of (\ref{eq:1d_shannon}) is obtained by $\tau_i^{(t)}+ s_i^{(t)}$.

\item When $0\leq s_i^{(t)}<-\tau_i^{(t)}$, as shown in Fig. \ref{fig:gsst_3}:
\begin{itemize}
\item For $x_i\geq0$, we have (\ref{eq:1d_shannon_1}). Since $\tau_i^{(t)}- s_i^{(t)}<0$, the $x_i$ that minimizes (\ref{eq:1d_shannon_1}) is $-\tau_i^{(t)}+ s_i^{(t)}$. 

\item For $x_i<0$, we have (\ref{eq:1d_shannon_2}). Since $-\tau_i^{(t)}- s_i^{(t)}>0$, the $x_i$ that minimizes (\ref{eq:1d_shannon_2}) is $\tau_i^{(t)}+ s_i^{(t)}$.
\end{itemize}
It's easy to verify that the minimum of (\ref{eq:1d_shannon_2}) is \emph{larger} than the minimum of (\ref{eq:1d_shannon_1}). Hence the global minimum of (\ref{eq:1d_shannon}) is obtained by $-\tau_i^{(t)}+ s_i^{(t)}$.

\item When $ s_i^{(t)}\geq-\tau_i^{(t)}$, as shown in Fig. \ref{fig:gsst_4}:
\begin{itemize}
\item For $x_i\geq0$, we have (\ref{eq:1d_shannon_1}). Since $\tau_i^{(t)}- s_i^{(t)}<0$, the $x_i$ that minimizes (\ref{eq:1d_shannon_1}) is $-\tau_i^{(t)}+ s_i^{(t)}$. 

\item For $x_i<0$, we have (\ref{eq:1d_shannon_2}). Since $-\tau_i^{(t)}- s_i^{(t)}<0$, the $x_i$ that minimizes (\ref{eq:1d_shannon_2}) is $0$.
\end{itemize}
(\ref{eq:1d_shannon}) is continuous at the point $x_i=0$. Hence the global minimum of (\ref{eq:1d_shannon}) is obtained by $x_i=-\tau_i^{(t)}+ s_i^{(t)}$.

\end{enumerate}
Combining the above $4$ different scenarios, we have the following results:
\begin{enumerate}
\item When $ s_i^{(t)}\geq0$, the solution to (\ref{eq:1d_shannon}) is $ s_i^{(t)}-\tau_i^{(t)}$.
\item When $ s_i^{(t)}<0$, the solution to (\ref{eq:1d_shannon}) is $ s_i^{(t)}+\tau_i^{(t)}$.
\end{enumerate}
This is exactly the soft shrinkage operator given in (\ref{eq:shrinkage}) on $ s_i^{(t)}$ with the threshold $\tau_i^{(t)}$.
\end{negative_threshold}

\section{First-Order Approximations of Other Regularization Functions}
\label{app:other_regularizers}

\begin{itemize}
\item $g_2(\vx)=\sum_{i=1}^N|x_i|^p$, where $\vx\in\mathbb{R}^N$. The function $g_2(\vx)$ is continuous in its domain, hence it is also lower semi-continuous. The first-order approximation in \eqref{eq:approx_shannon_entropy} is updated accordingly:
\begin{align}
\begin{split}
g_2(\vx)&\approx \|\vx^{(t)}\|_p^p+\left<|\vx|-|\vx^{(t)}|, \nabla g_2(|\vx^{(t)}|)\right> \,,
\end{split}
\end{align}
where $0<p<1$. The gradient with respect to $|x_i|$ is:
\begin{align}
\frac{\partial g_2(\vx)}{\partial |x_i|} = p|x_i|^{p-1}\,.
\end{align}

\item $g_4(\vx)=\frac{\|\vx\|_1}{N\cdot\max_j |x_j|}-1$, where $\vx\in\mathbb{R}^N$ and $\vx\neq\boldsymbol 0$. The function $g_4(\vx)$ is continuous in its domain. Hence it is also lower semi-continuous.

Since $g_4(\vx)$ is not differentiable with respect to the modulus $|\vx|$, we use its subderivative to get the first order approximation:
\begin{align}
\begin{split}
g_4(\vx)&\approx g_4(|\vx^{(t)}|)+\left<|\vx|-|\vx^{(t)}|, \partial g_4(|\vx^{(t)}|)\right>\,.
\end{split}
\end{align}

The subderivative with respect to $|x_i|$ is:
\begin{align}
\frac{\partial g_4(\vx)}{\partial |x_i|} = \frac{1}{N}\left(\frac{1}{\|\vx\|_\infty}-\delta_\infty(i)\frac{\|\vx\|_1}{\|\vx\|_\infty^2}\right)\,,
\end{align}
where $\delta_\infty(i)$ is given in \eqref{eq:delta_infty}.

\end{itemize}

\end{appendices}


\ifCLASSOPTIONcaptionsoff
  \newpage
\fi



%
\bibliographystyle{IEEEbib}
\bibliography{refs}

\end{document}